\documentclass[
reprint,
amsmath,amssymb,
aps,
prl,
]{revtex4-2}

\usepackage{graphicx}% Include figure files
\usepackage{dcolumn}% Align table columns on decimal point
\usepackage{bm}% bold math
\usepackage{hyperref}% add hypertext capabilities
\usepackage{graphicx} % Required for inserting images
\usepackage{xcolor}
\usepackage{makecell}
\usepackage{amsmath}
\usepackage{soul}
\usepackage{orcidlink}
\usepackage[dvipsnames]{xcolor}

\mathchardef\mhyphen="2D

\begin{document}

\title[Article Title]{Learning Hamiltonians for solid-state quantum simulators}

\author{Jaros\l{}aw Paw\l{}owski\orcidlink{0000-0003-3638-3966}}
\email{jaroslaw.pawlowski@pwr.edu.pl}
\affiliation{Institute of Theoretical Physics, 
%Faculty of Fundamental Problems of Technology,
Wroc{\l}aw University of Science and Technology,
Wybrze\.{z}e Wyspia\'{n}skiego 27,
50-370 Wroc{\l}aw, Poland}

\author{Mateusz Krawczyk\orcidlink{https://orcid.org/0000-0003-0466-814X}}
\affiliation{Institute of Theoretical Physics, 
%Faculty of Fundamental Problems of Technology, 
Wroc{\l}aw University of Science and Technology,
Wybrze\.{z}e Wyspia\'{n}skiego 27,
50-370 Wroc{\l}aw, Poland}

\begin{abstract}
%We introduce a generalizable framework for learning to identify effective Hamiltonians directly from experimental data in solid-state quantum systems. Our approach is based on a physics-informed neural network architecture that embeds physical constraints directly into the model structure. Unlike purely data-driven supervised schemes, the proposed unsupervised autoencoder-based method incorporates the governing physics (here, the S-matrix formalism) within the decoder network, ensuring that the learned representations remain physically meaningful. Through numerical learning experiments, we demonstrate automated characterization of programmable solid-state simulators from transport measurements, exemplified by a triple quantum dot chain. The trained model generalizes beyond the training domain and accurately infers Hamiltonian parameters from transport data. While the model has finite capacity---leading to degraded performance when the parameter space becomes excessively large or structurally diverse---we identify regimes in which robust generalization is maintained. We further show how to train the model to handle noisy measurements, reflecting realistic experimental conditions and also prove that the model can be successfully trained on transport data generated outside the Hamiltonian family assumed by the physics decoder, while still identifying meaningful effective Hamiltonian parameters.
We introduce a generalizable framework for identifying effective Hamiltonians directly from experimental data in solid-state quantum systems. Our unsupervised autoencoder-based approach incorporates the governing physics (here, the S-matrix formalism) directly into the decoder, enabling physically meaningful Hamiltonian inference without labeled Hamiltonian parameters. Through numerical experiments on a triple quantum dot chain, we demonstrate automated characterization of programmable solid-state simulators from transport measurements. The model accurately infers Hamiltonian parameters and generalizes beyond the training domain, while exhibiting finite capacity as the parameter space becomes increasingly broad or diverse. We further demonstrate robustness to noisy measurements and show that the model can be trained on transport data generated outside the Hamiltonian family assumed by the physics decoder, while still identifying meaningful effective parameters.
\end{abstract}

\maketitle

\section{Introduction}
Machine learning (ML) techniques have been widely explored in solid-state physics, including phase identification~\cite{Nieuwenburg2017,Chng2017,Zhang2018,Yunwei2020} and materials discovery through the prediction of chemical and electronic properties~\cite{Schutt2018,Merchant2023,Cheetham2024,Qi2025}. A central challenge in this area is the extraction of effective Hamiltonians 
$\mathcal{H}$ from experimental or simulated observables~\cite{Gebhart2023}. Existing approaches broadly fall into two categories: optimization-based inverse methods, such as evolutionary algorithms and heuristic searches~\cite{Lunczer2019,Lew2023,Inui2024,Thamm2024}, and supervised learning schemes~\cite{Lupi2025,Karjalainen2023,Feng2024,grepkowa2023,Wang2017,Taylor2024disorderlearning}, which often struggle with robustness and generalization.

Hamiltonian learning (HL) has shown particular promise in nanoscale solid-state platforms, including the identification of quantum nanomagnets from spectral data~\cite{Lupi2025,Karjalainen2023}, classification of skyrmionic magnetic textures~\cite{Feng2024}, and ML-assisted control and readout of molecular spin qubits~\cite{Bonizzoni2022}. Beyond transport measurements, HL has also been applied to density-of-states data for automated band-structure inference~\cite{Henderson2023} and to the analysis of local patterns in moir\'e materials~\cite{Khosravian_2024,liu2025}. Learning tight-binding Hamiltonians is likewise emerging as an important research direction~\cite{Gu2024,Choudhary2025}. Finally, learning Hamiltonian dynamics constitutes another active line of research, with applications ranging from classical systems~\cite{Mattheakis2022,greydanus2019hamiltonian} and generic quantum dynamics~\cite{Mirani2024,StilckFranca2024} to superconducting quantum processors~\cite{Hangleiter2024}.

Gate-defined quantum dots (QDs) hosting single electrons or holes are promising platforms for solid-state quantum computing due to their electrical tunability and scalability, and have recently attracted interest as programmable quantum simulators~\cite{Borsoi2024,Mills2019,Shandilya2025,hernandes2026}. In particular, minimal realizations of the Kitaev chain in QD arrays---based on elastic cotunneling and crossed Andreev reflection and supporting so-called \emph{poor man’s} Majorana zero modes (MZMs)~\cite{Leijnse2012}---have attracted significant attention~\cite{Wang2023,Bordin2025}. Rashba quantum dot chains have also been proposed as alternative simulator platforms capable of hosting MZMs~\cite{Fulga2013,Maska2017}.

At the same time, ML-assisted autotuning of QD-based quantum simulators using transport measurements is gaining increasing interest~\cite{Darulov2020,Lunczer2019,Zwolak2023,roux2025,losert2025,hernandes2026}. The use of transport measurements in the form of \textit{conductance maps}~\cite{Donarini2024,Shandilya2025} to extract information about the system Hamiltonian~\cite{Blonder1982,grepkowa2023,Wang2017,Gebhart2023,Thamm2024,Taylor2024disorderlearning,Taylor2025analysis} appears to be a promising route toward automated parameter tuning. In particular, inverting measured conductance matrices to determine electrostatic potential disorder---using evolutionary optimization~\cite{Thamm2024}, supervised neural networks~\cite{Taylor2024disorderlearning,vandriel2024,Taylor2025analysis}, or hybrid approaches~\cite{Taylor2025disordermitigation}---provides a natural starting point for precise inference of $\mathcal{H}$ parameters relevant to MZMs.

\begin{figure*}[tb]
	\centering
	\includegraphics[width=0.99\linewidth]{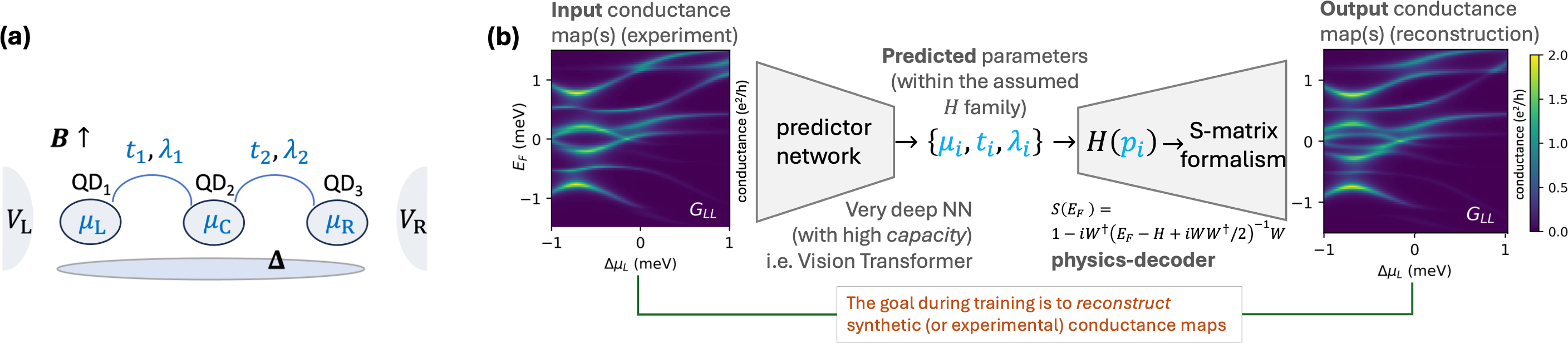}
	\caption{(a) Chain of the Rashba QDs. (b) Proposed HL architecture composed of encoder---predictor and physics-decoder.}
	\label{fig:scheme}
\end{figure*}

Physical constraints can be incorporated into neural network (NN) models in two principal ways: by enforcing symmetry-respecting representations (geometric deep learning~\cite{bronstein2021}) or by constraining solutions to satisfy governing differential equations, as in physics-informed neural networks (PINNs)~\cite{Raissi2019,Kovachki2023,Li2024}. Both approaches typically rely on supervised training, which can be limiting in realistic HL scenarios where the parameters of effective Hamiltonians are not directly accessible. Here, we adopt a physics-decoder (PD) architecture~\cite{Krawczyk2024a,Kliczkowski2024,Leszczynski2024,krawczyk2026}, shown in Fig.~\ref{fig:scheme}(b), inspired by PINNs but formulated as a fully unsupervised autoencoder-based framework.
We employ this paradigm specifically for unsupervised Hamiltonian identification from transport measurements, with the central goal of learning a generalizable inverse map from experimentally accessible observables to effective microscopic Hamiltonian parameters.
In this scheme, the encoder infers Hamiltonian parameters, while the decoder explicitly implements the underlying physics for the assumed Hamiltonian family---a strategy that can be readily generalized beyond the present setting. HL with physics-informed training schemes remains relatively underexplored and constitutes an active area of research~\cite{Elhamod2022,Gu2024,Li2022,Choudhary2025}.

\section{Quantum dots chain}
To demonstrate the working principle of the method, we define a simple simulator consisting of a chain of $N=3$ QDs, yet sufficiently complex to illustrate the robustness of the approach. We consider a $\mathcal{H}$ family that defines the Kitaev-chain simulator based on a Rashba QD chain coupled with a nearby $s$-wave superconductor~\cite{Fulga2013}, depicted in Fig.~\ref{fig:scheme}(a).
The Hamiltonian describing chain of spinful single-level QDs with local potential $\mu_n$, Zeeman energy $V_\mathrm{Z}$, proximity-induced superconducting ($s$-wave) pairing \(\Delta_n\), and Rashba spin-orbit vector $\boldsymbol{\lambda_n}$, which modifies the inter-dot hopping $t_n$, is defined as:
\begin{align}
H^\mathrm{QDs} = {} &
\sum_{n,s,s'}
\Bigl[
(-\mu_n \sigma_0 + V_\mathrm{Z} \sigma_z)_{ss'} \, c_{n,s}^\dagger c_{n,s'}
\Bigr. \nonumber \\
& + \tfrac{1}{2} 
\!\left(
\left(\Delta_{n} i\sigma_y \right)_{ss'} 
c_{n,s}^\dagger c_{n,s'}^\dagger 
+ \mathrm{h.c.} 
\right)\nonumber \\
& +\Bigl.\left(
t_n (e^{i\boldsymbol{\lambda}_n \cdot \boldsymbol{\sigma}})_{ss'} 
c_{n,s}^\dagger c_{n+1,s'} 
+ \mathrm{h.c.}
\right)
\Bigr],
\label{eq:qds}
\end{align}
where $\sigma_i$ are Pauli matrices in spin space (indexed by $s$ and $s'$), and $n$ numbers QDs in the chain. For simplicity we assume uniform pairing $\Delta_n=0.25$~meV, the Rashba vector in a form $\boldsymbol\lambda_n=\lambda_n[0,1,0]$, and the global Zeeman field $V_\mathrm{Z}=0.5$~meV.
The rest control parameters, i.e., $\{\mu_n$, $t_n$, $\lambda_n\}\equiv P$ (7 in total for $N=3$ QDs chain) can be tuned electrically (via local gating). One set (yet uniform) of parameters includes: $\mu_n=0.6\,\mathrm{meV}$, $t_n=0.25\,\mathrm{meV}$, $\lambda_n=0.27\,\pi$ we call \textit{reference parameters}, $P_0$.
They define so-called \emph{sweet spot} leading to MZMs emergence, discussed in detail in Appendix~A.
The configuration of eigenstates for $N=7$ and $N=3$ is shown in Fig.~\ref{fig:mzm}(a) and Figs.~\ref{fig:mzm}(b,c) respectively.
The Rashba length $\lambda_n = 0.27\,\pi$ was tuned so that at $\mu = 0.6$~meV two energy levels touch at zero energy (c.f. Fig.~\ref{fig:mzm}(b,c)).
%In Figs.~\ref{fig:mzm}(b,c) there are presented eigenstates as a function of voltage offset for the system of triple-QD Rashba chain. Colors denote: (b) edge dots (L and R) occupation, and (c) particle-hole symmetry, showing the region where Majorana zero modes emerge (pointed by orange arrow).

\section{Transport measurements}
\begin{figure}[bt]
	\centering
	\includegraphics[width=1.\columnwidth]{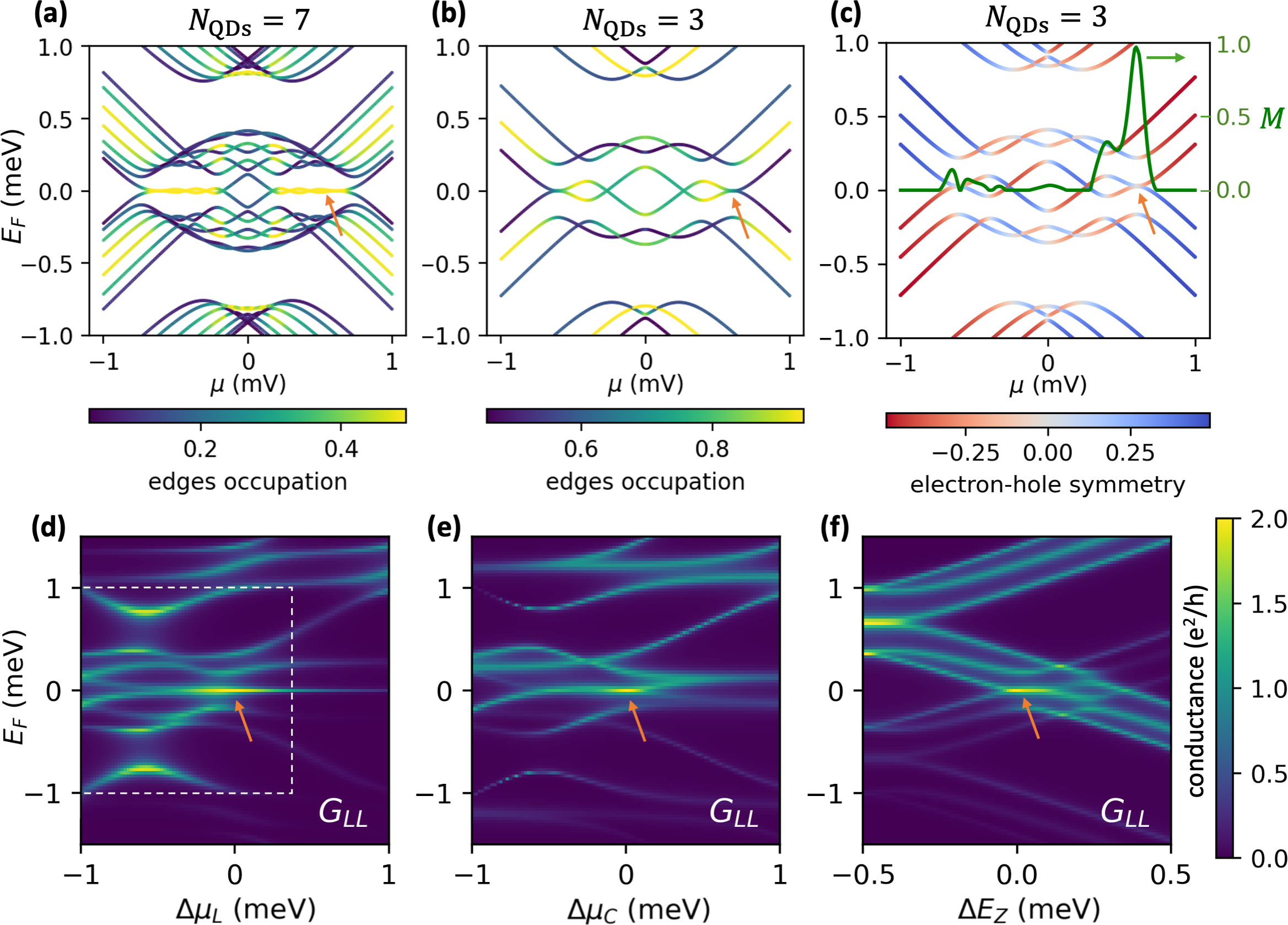}
	\caption{Rashba chain of QDs proximitized by an $s$-wave superconductor. Top row: eigenstates as a function of voltage offset with colors encoding: (a,b) edges ($L$ and $R$ dot) occupation, (c) electron-hole symmetry. Reference parameters configuration $P_0$, giving MZM-like states, is marked by orange arrow. Bottom row: conductance maps as a function of (d) voltage applied to the left dot, (e) voltage on the center dot, and (f) the Zeeman field.}
	\label{fig:mzm}
\end{figure}

The analyzed system can be characterized via transport measurements. The conductance $G$ through the QD chain is calculated using the $S$-matrix formalism in the wide-band limit~\cite{Donarini2024} via the Weidenm\"uller formula~\cite{Bordin2025,christiansen2009} for QDs coupled to normal leads:
\begin{equation}\label{eq:smatrix}
    S[H](E_F)=\mathbf{1}-iW^\dagger(E_F-H+\frac{i}{2}WW^\dagger)^{-1}W,
\end{equation}
with the tunneling matrix defined as $W=\mathrm{diag}(1,0,1)\otimes\mathrm{diag}(\sqrt{\Gamma},\sqrt{\Gamma},-\sqrt{\Gamma},-\sqrt{\Gamma})$ and the dot-lead coupling $\Gamma=0.1$~meV. 
If we reshape the $S$-matrix accordingly: $S=S_{n,p,s,n'\!,p'\!,s'}$ ($n=L,C,R$ indexing dots, $p=1,2$ particle, and $s$---spin subspaces), then the respective reflection matrices: $s^{ee}_{s,s'}\!(i,j)=S_{i,1,s,j,1,s'}$ and $s^{he}_{s,s'}\!(i,j)=S_{i,2,s,j,1,s'}$ give differential conductance as,
\begin{align}\label{eq:gdef}
G[S(E_F)]_{ij}=&\frac{\mathrm{d}I_i}{\mathrm{d} V_j}=2\delta_{ij}-\mathrm{tr}\!\left(s^{ee}(i,j)s^{ee}(i,j)^\dagger\right)\nonumber\\
+&\,\mathrm{tr}\!\left(s^{he}(i,j)s^{he}(i,j)^\dagger\right)
\end{align}
in unit of $e^2\!/h$, $i,j=L,R$ denoting left ($L$) or right ($R$) lead---see Fig.~\ref{fig:scheme}(a), and $E_F$ being the Fermi energy in the leads. Note that the same formula is implemented in the PD shown in Fig.~\ref{fig:scheme}(b).
The input maps include 4 conductance $G_{ij}$ components: $G_{LL}$, $G_{LR}$, $G_{RL}$, and $G_{RR}$, where for instance $G_{LL}=\frac{dI_{L}}{dV_{L}}$, with $I_{L}$ denoting current through the left lead, and $V_{L}$ is the bias voltage of the left lead. Similarly, other components can be defined by using different combinations of the leads. 
Notably, each conductance map is defined by $100\times100$ 2D plot of $G_{ij}$ as a function of some parameter and $E_F$.
We utilize 4 maps per each component $G(E_f,\Delta l)$: 3 for \(l=\mu_{n=L,C,R}\) variation (with respect to actual parameter $P$) and 1 for \(l=V_\mathrm{Z}\) variation: 16 maps in total serving as the input. Figs.~\ref{fig:mzm}(d-f) shows the conductance maps ($G_{LL}$ component) for the reference parameters $P_0$, highlighting the emerged zero-bias peak (orange arrow).

\section{Neural architecture and training}
The proposed method uses the conductance maps tensor $G$ as a high-dimensional visual input for the predictor NN (encoder), as well as to define the physics-inspired decoder---PD (teacher network) for unsupervised training. The model architecture is presented in Fig.~\ref{fig:scheme}(b). It is of autoencoder-form with predictor NN coupled to PD network through parameters $P$ \emph{latent} space, and can be denoted as
\begin{equation}
    \hat{G}=\mathrm{PD}(\mathrm{NN}(G)=P),
    \quad\mathrm{PD}(P)\equiv G[S[\mathcal{H}(P)]].
\end{equation}
The encoder analyzes the $G$ maps and predicts the parameter vector $P$; PD explicitly implements the transport physics using Eqs.~\ref{eq:smatrix} and \ref{eq:gdef}, 
with the assumed systems family $\mathcal{H}=H^\mathrm{QDs}$, defined in Eq.~\ref{eq:qds}. The objective is for the reconstructed map $\hat{G}$ (the PD output) to match the (encoder) input $G$---by minimizing loss function $\mathcal{L}(G,\hat{G})=||G-\hat{G}||^2$.
We adapt a pretrained vision transformer (ViT) architecture~\cite{dosovitskiy2021} (patch size $16\times16$, 12 heads, 12 layers, $\sim86$M~parameters) with input layers adjusted to process 16-channel $(16\times100\times 100)$ input tensor $G$ and returning vector of the Hamiltonian parameters $P$. Training ($\mathcal{L}$ minimization) is performed for 300 epochs (with a properly tuned learning-rate scheduler) using 10-50k data points ($G(P_\mathrm{in})$ maps) for parameter vectors $P_\mathrm{in}$ sampled from different regions of the parameter space, as described in the Results section.

\section{Results}
\begin{figure}[tpb]
	\centering
	\includegraphics[width=0.99\columnwidth]{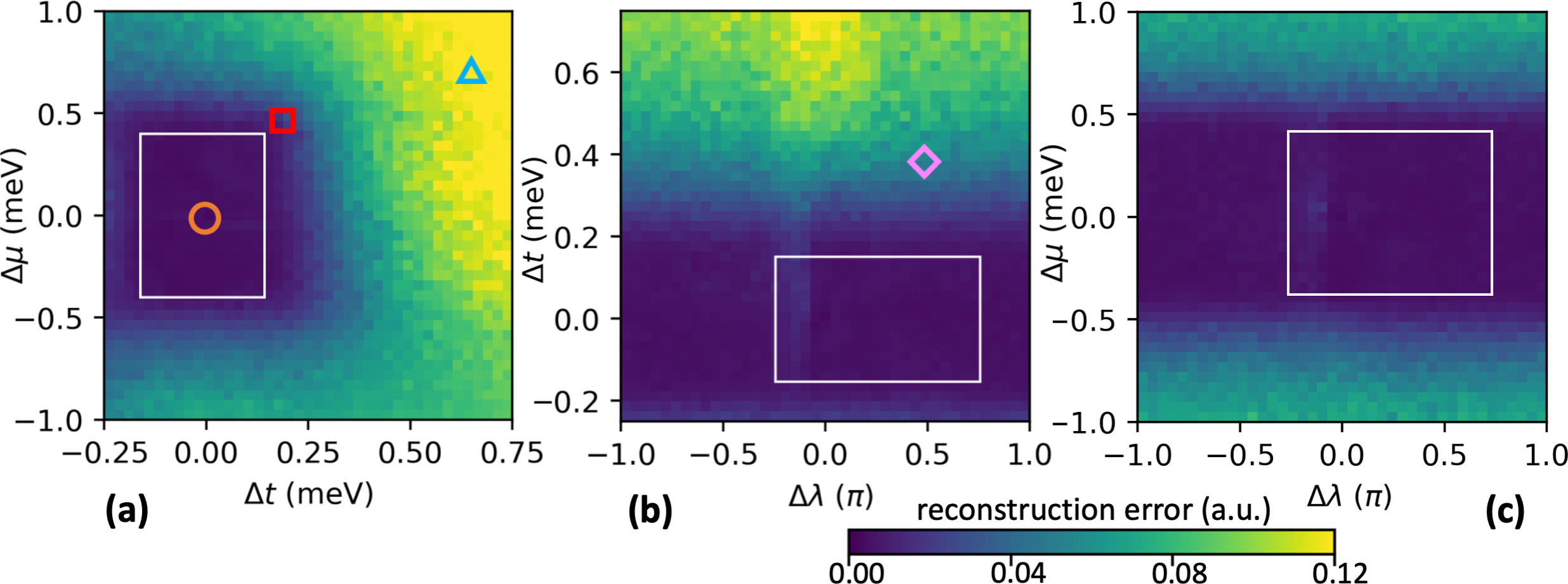}
    \includegraphics[width=0.97\columnwidth]{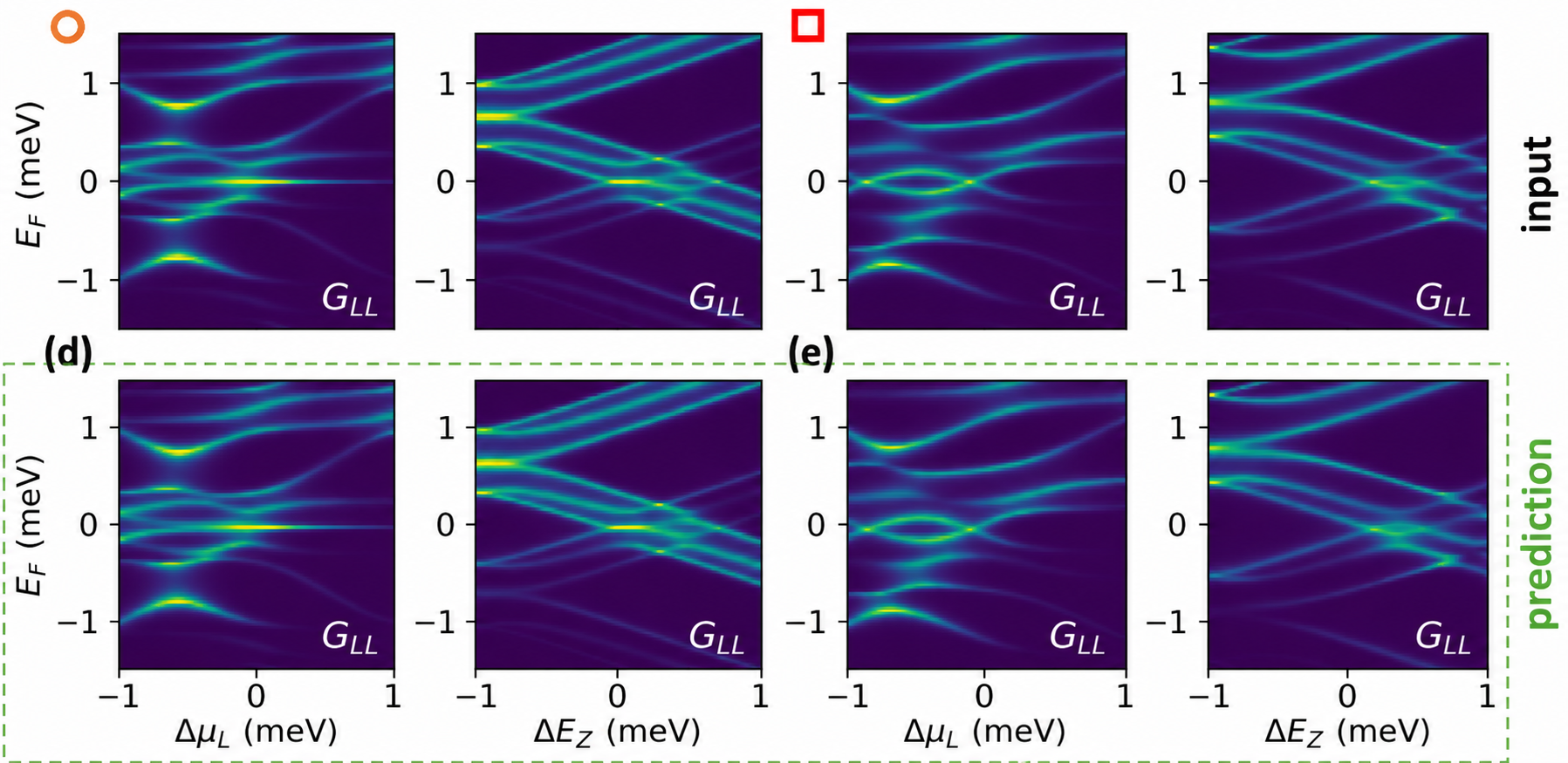}
    \includegraphics[width=0.97\columnwidth]{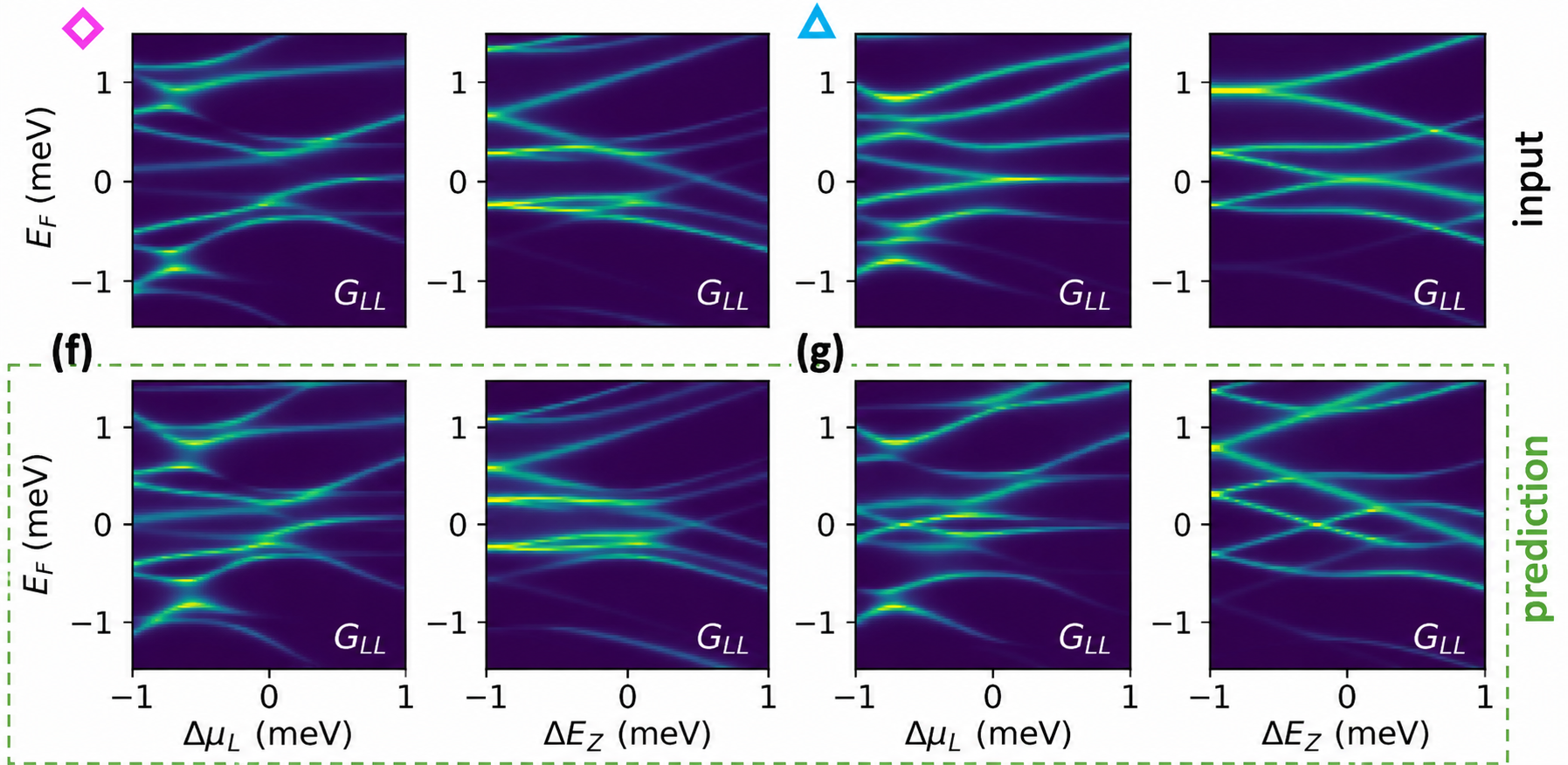}
	\caption{Results for the HL of the Rashba chain, Eq.~\ref{eq:qds}, of three quantum dots proximitized by an $s$-wave superconductor. Top row (a-c): error for the conductance maps reconstruction (via the predicted parameters) as a function of parameter variations---the white box denotes the region where training data was sampled. Middle (d,e) and bottom (f,g) rows: example input and predicted conductance maps (shown as middle/bottom pairs) for various parameter configurations indicated by the respective symbols. The default configuration that yields MZMs is marked with an orange circle}
	\label{fig:rashba_hl}
\end{figure}

\begin{figure}[tpb]
	\centering
	\includegraphics[width=0.99\columnwidth]{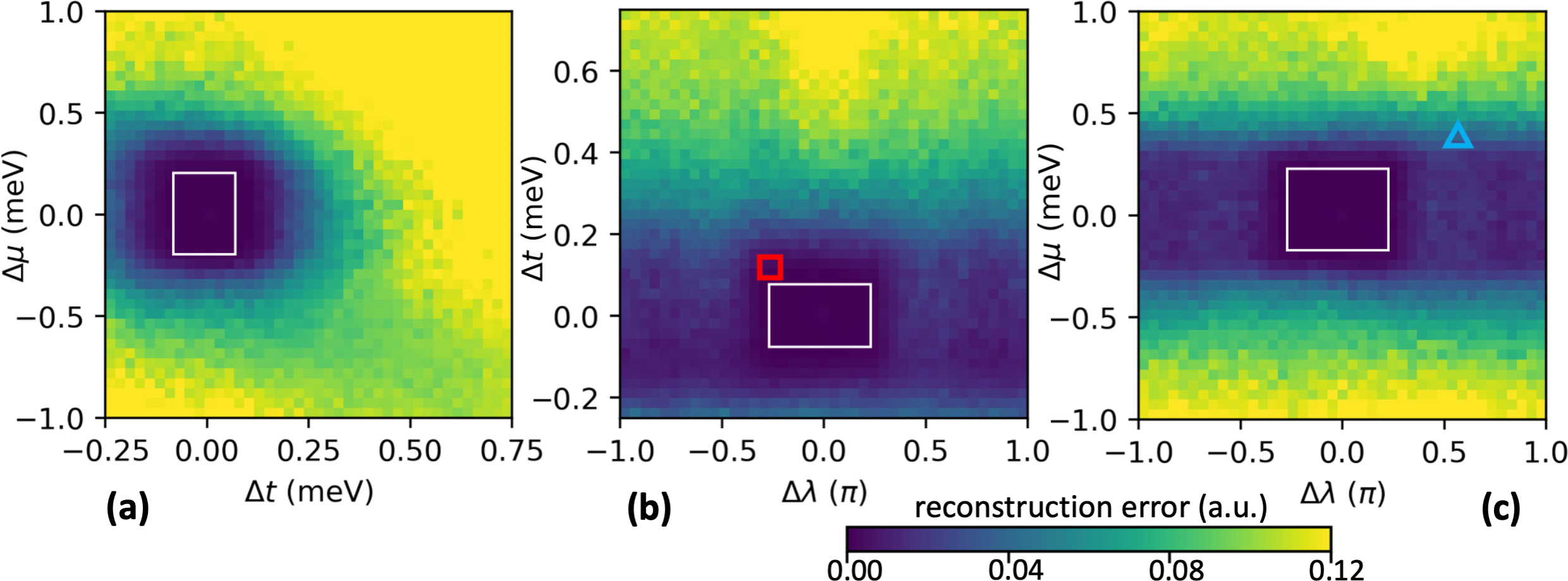}
    \includegraphics[width=0.97\columnwidth]{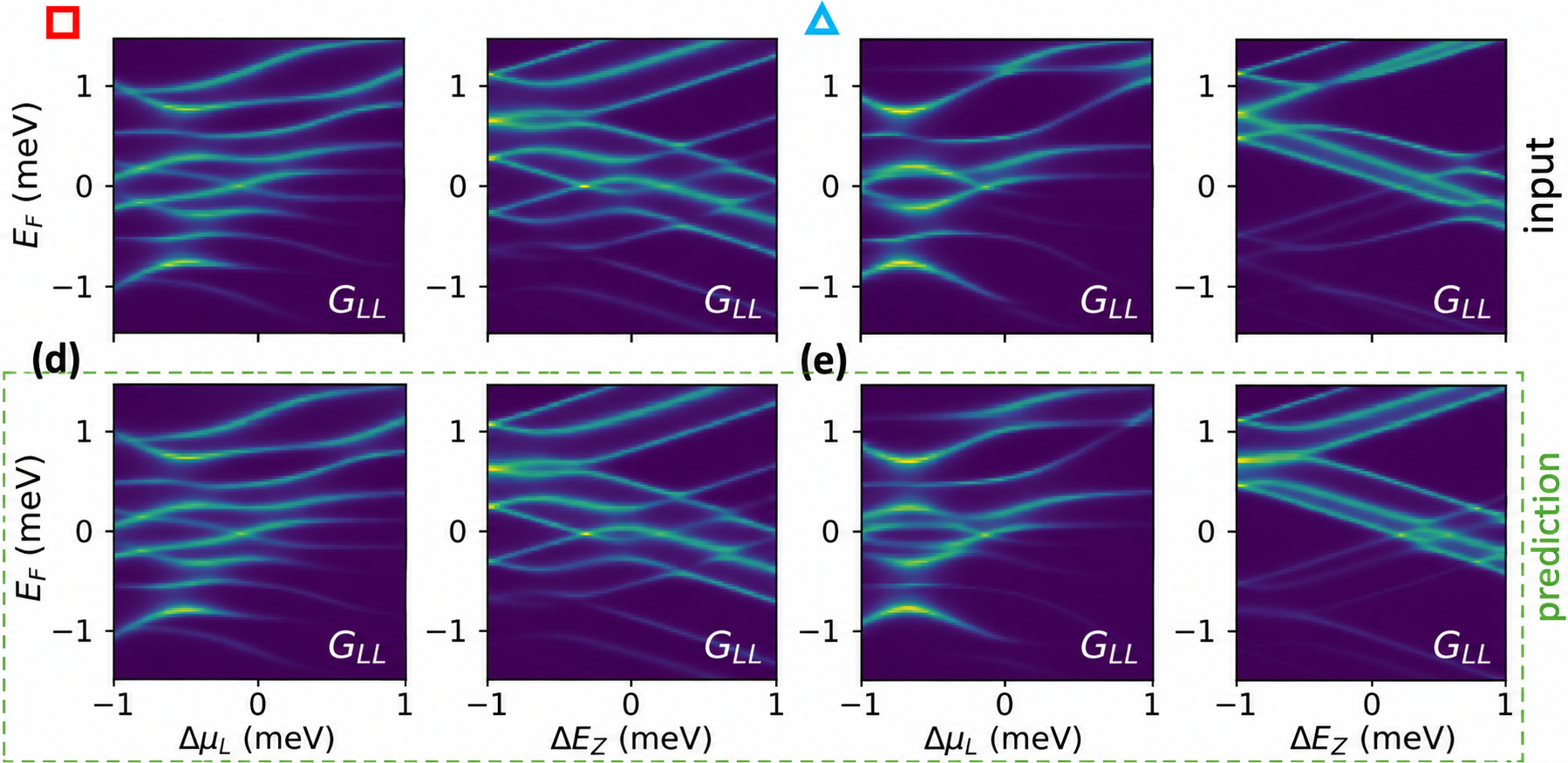}
	\caption{HL results for a smaller training range, denoted by white boxes on the conductance reconstruction error maps (a-c) within the parameter space. Two specific $G$ reconstruction cases (d,e) are indicated in the parameter space by a red square in (b) and a blue triangle in (c), respectively.}
	\label{fig:smaller_range}
\end{figure}

\begin{figure}[t]
	\centering
	\includegraphics[width=0.99\columnwidth]{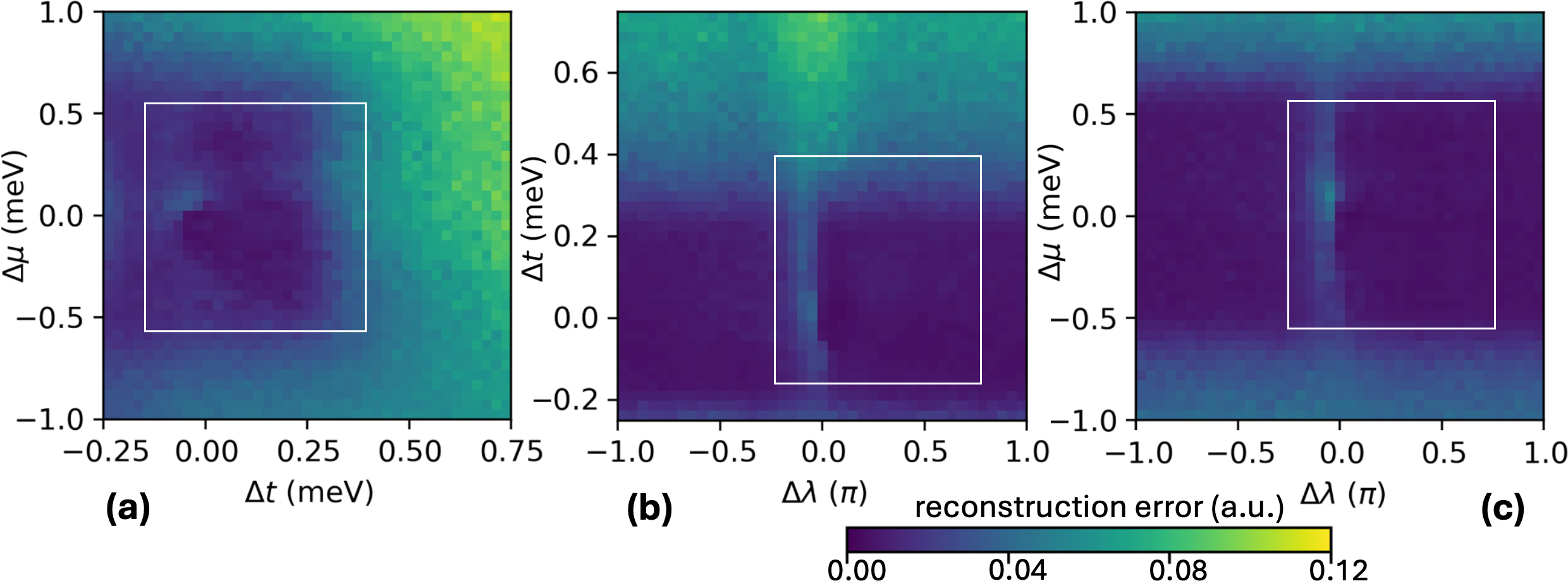}
	\caption{HL results for a larger training range, indicated by wider white boxes on the conductance reconstruction error maps (a-c). In this case, 50k training samples were used.}
	\label{fig:larger_range}
\end{figure}

To demonstrate the validity of the PD concept, we performed HL of $\mathcal{H}=H^\mathrm{QDs}$ parameters, i.e. $P=\{\mu_\mathrm{L}, \mu_\mathrm{C}, \mu_\mathrm{R}, t_1, t_2, \lambda_1, \lambda_2\}$ for the triple-QD Rashba chain, as described above. After the training using synthetic $G(P_\mathrm{in})$ maps with $P_\mathrm{in}$ sampled from training sets denoted as white boxes in Figs.~\ref{fig:rashba_hl}-\ref{fig:noise}, the predictor NN can predict the (learned) Hamiltonian parameters $P$ for a given $G$.
Sampling $\Delta l$ in the range $[l_a,l_b]$ means that parameter $l\in [l_a,l_b]+P_{0,l}$, e.g., all $\mu_n$ were sampled from some $[\mu_a,\mu_b]+0.6$~meV.
The robustness of the predictor is shown in Fig.~\ref{fig:rashba_hl}. 
The middle and bottom rows (d-g) show pairs of: input conductance maps (middle), and conductance maps \textit{reconstructed} (calculated) using the predicted parameters (bottom).
The better the parameter prediction, the closer the maps in each pair.
The pairs presented are for different parameter configurations, indicated by the respective symbols in the top row (a-c).
In the top row the reconstruction error $\langle|G-\hat{G}|\rangle$ is presented as a function of various parameters variation $(\Delta l_1,\Delta l_2)$ giving three different cuts in the parameter $P$ space.
Reconstruction error is averaged over all channels and pixels on $G_{ij}$ maps, and over variations of different parameters of a given category (e.g., $\Delta\mu$ means all variations of potentials $\Delta\mu_n$).

Inspection of the $G_{LL}$ maps in Fig.~\ref{fig:rashba_hl} for parameters (d) within the training set (denoted by an orange circle on (a), corresponding to $P_0$), or (e) close to it (red square) shows that the reconstructed maps match the input maps almost perfectly. Moving slightly further away (f, pink diamond), small reconstruction errors begin to appear, although the overall structure of the conductance bands remains qualitatively well reproduced. However, when we move farther from the training set (g, blue triangle), the resulting structure of the conductance peaks in the $G$ maps becomes noticeably different.
These results demonstrate that the PD architecture is capable of learning the complex structure of the conductance features that arise even in this simple triple-QD system. More importantly, the model provides accurate predictions for regions of parameter space outside the training domain (outside the white rectangles in Fig.~\ref{fig:rashba_hl}(a-c). This indicates that the physics-informed PD architecture possesses a strong ability to generalize.
Importantly, the sampling of the 7-dimensional training space is relatively sparse: 10k samples correspond to approximately 
$(10^4)^\frac{1}{7}\simeq3.7$ points per dimension.

To better investigate this property, we evaluated the NN predictor performance for a smaller training region---indicated in Fig.~\ref{fig:smaller_range}, and for a larger one---shown in Fig.~\ref{fig:larger_range}. As expected, the smaller training region (reduced by a factor of $2^3$) leads to accurate parameter predictions only within a correspondingly smaller volume of the parameter space $P$. Nevertheless, even in this case there remains a region outside the training set where the model generalizes well. Moreover, the ratio of this generalization volume to the training volume is slightly larger than in the configuration shown in Fig.~\ref{fig:rashba_hl}. This can be attributed to the fact that, in the vicinity of the reference parameters $P_0$ (emergence of MZMs), the structure of peaks on the $G$ map is relatively similar, making it easier for the NN to learn.

In the opposite extreme, presented in Fig.~\ref{fig:larger_range}---we observe that for a larger training volume (increased by a factor of 
$1.5\times2\times1=3$) the model correctly predicts the parameters 
$\mathcal{H}(P)$ over nearly the entire explored parameter space $P$. Importantly, in this case we increased the number of training samples to 50k (in the previous configurations shown in Figs.~\ref{fig:rashba_hl} and \ref{fig:smaller_range}, we used 10k samples).
However, as a characteristic trade-off, the price of covering such a broad range of $G$-maps variability is a slightly reduced average accuracy of parameter predictions, even within the training region itself. This clearly indicates that we are approaching the NN capacity limit.

It is also important to note that the inverse mapping from $G$ maps to $\mathcal{H}$ parameters $P$ is not guaranteed to be globally unique. In particular, symmetries and gauge freedoms of the Hamiltonian may lead to different parameterizations producing identical or nearly identical transport observables $G$. In our case, an example is the transformation: $t\rightarrow-t$ and $\lambda\rightarrow\lambda+\pi$. In the present approach, such ambiguities are mitigated by restricting the predictor output to physically and experimentally relevant parameter ranges $[l_a,l_b]+P_{0,l}$, thereby selecting a particular branch of the inverse problem. More generally, this restriction could be relaxed by employing a multi-head encoder that predicts several candidate parameter sets, allowing the network to explicitly represent multiple branches of a non-unique inverse map. Such a generalization would require further investigation; here, for simplicity, we restrict the predicted parameters to predefined physically relevant ranges.

\section{Resilience to noise}

In the proposed approach, we assume that the NN model is first trained on synthetically generated data and subsequently transferred to experimental measurement data. However, experimental data may exhibit characteristics different from those of the synthetic dataset; in particular, they may contain noise.

\subsection{Noise in conductance}
\begin{figure}[tb]
	\centering
	\includegraphics[width=0.99\columnwidth]{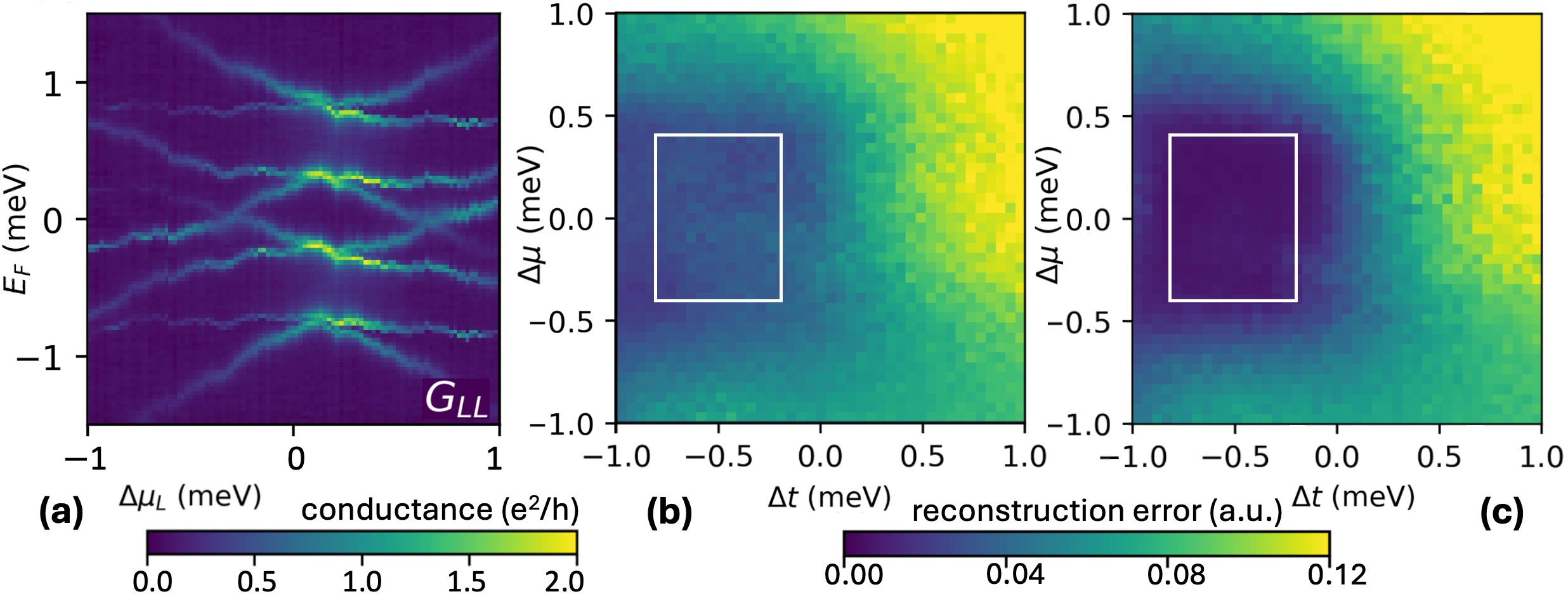}
	\caption{HL from noisy data: (a) noisy conductance maps lead to (b) slightly degraded predictions. However, retraining the model with noisy samples included in the training data (c) restores the prediction performance.}
	\label{fig:noise}
\end{figure}
To test this and emulate realistic experimental conditions, we augmented the synthetic conductance maps $G_{ij}(E_f,\Delta l)$ by including two different noise contributions, according to
\begin{equation}
\tilde{G}_{ij}(E_f,\Delta l)=G_{ij}^{\mathrm{warp}}(E_f,\Delta l)
+\delta G_{ij}^{\mathrm{add}}(E_f,\Delta l),
\end{equation}
where $E_F$ is the Fermi energy in the leads and $l$ denotes the detuned control parameter, corresponding either to the chemical potential of a selected quantum dot ($l=\mu_n$) or to the Zeeman field ($l=V_Z$). The term $G_{ij}^{\mathrm{warp}}$ describes distortions caused by charge noise and gate drift, while $\delta G_{ij}^{\mathrm{add}}$ accounts for additive instrumental noise.
Both are discussed in detail in Appendix~B. 

We next examine the robustness of the proposed method to noise defined in this way---an example of a noisy conductance map 
$\tilde{G}$ is shown in Fig.~\ref{fig:noise}(a). When testing the model presented in Fig.~\ref{fig:rashba_hl} on such noisy maps, the prediction quality naturally degrades---see Fig.~\ref{fig:noise}(b). However, if the noise present in the experimental setup is well characterized, we can sample $\tilde{G}$ and add to the training data (10k samples were added).
In Fig.~\ref{fig:noise}(c) we show the reconstruction error for the NN model trained on the augmented conductance map set. It is visible that the model is now capable of predicting the $\mathcal{H}$ parameters even from the noisy conductance maps.

Note also that the PD architecture can be used for efficient GPU-parallelized synthetic data generation, as was done in this work for conductance map sets.

\subsection{Hamiltonian-level disorder}

\begin{figure}[b]
	\centering
    \includegraphics[width=0.89\columnwidth]{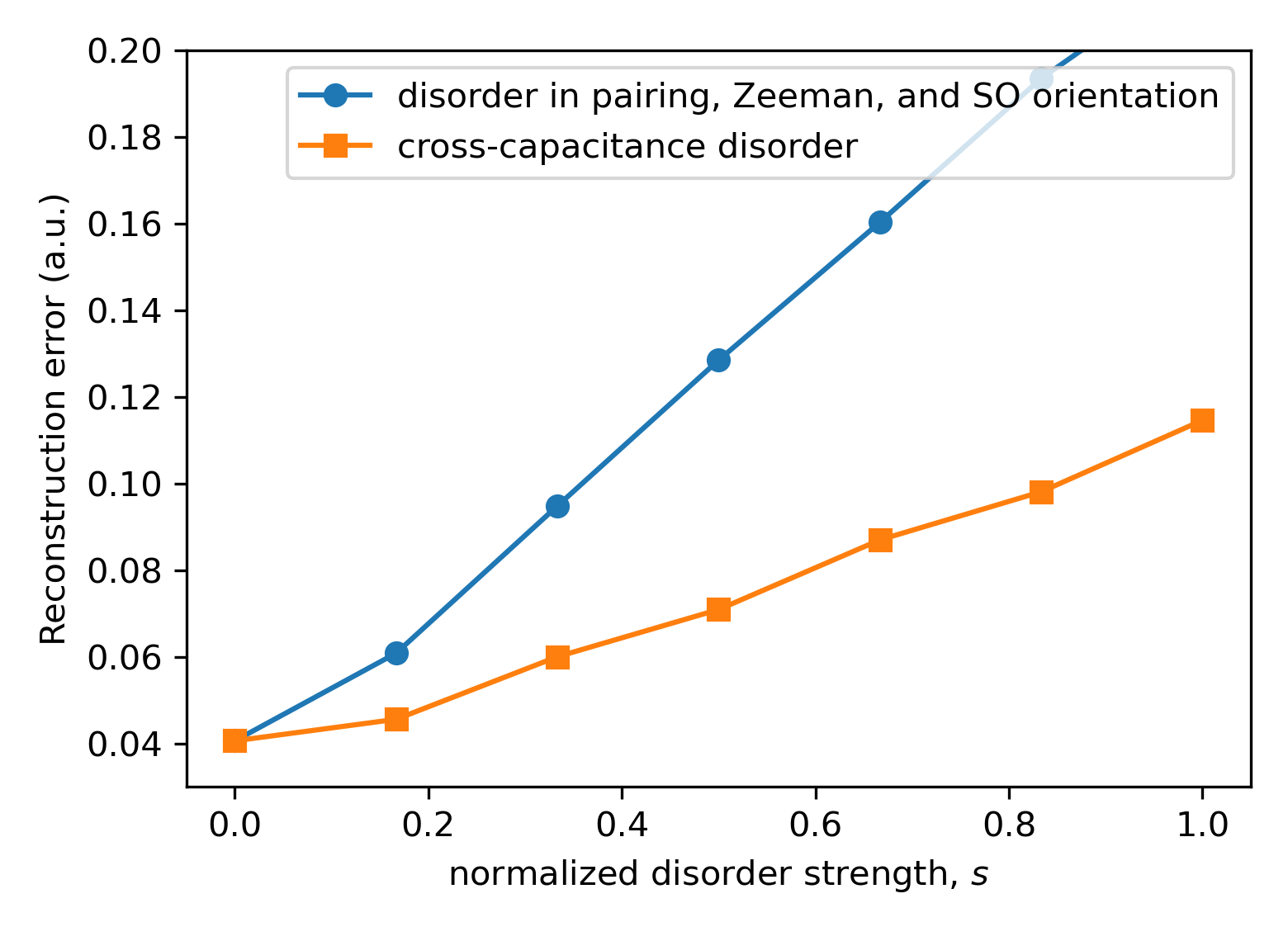}
	\caption{Prediction robustness against disorder in Hamiltonian parameters outside the predicted parameter space $P$ for two cases: (blue) local disorder in the pairing, Zeeman, and Rashba orientation terms; (orange) cross-capacitance disorder of the QDs. The disorder strength $s$ is normalized to the range $[0,1]$, with its detailed definition provided in Appendix~C.}
	\label{fig:diss}
\end{figure}

Having established the robustness of the model against noise in the conductance maps, a natural next step is to examine its sensitivity to imperfections at the Hamiltonian level. This distinction is important because measurement noise modifies the observed signal, whereas Hamiltonian-level disorder introduces a mismatch between the physical system generating the data and the idealized Hamiltonian family assumed by the physics decoder. We consider two representative sources of such model mismatch. First, we introduce static sample-to-sample variations of parameters that are fixed in the decoder, namely the proximity-induced pairing $\Delta_n$, the local Zeeman field $V_{\mathrm{Z},n}$, and the orientation of the Rashba spin-orbit field $\hat{\boldsymbol{\lambda}}_n$. These perturbations are kept fixed during the acquisition of each complete set of conductance maps and therefore represent device-level Hamiltonian disorder rather than measurement noise. Second, we account for cross-capacitance between electrostatic gates. Instead of assuming that detuning gate $k$ modifies only the corresponding local potential $\mu_k$, we introduce a capacitance matrix $\mathbf{A}$ such that $\delta\boldsymbol{\mu}=\mathbf{A}\mathbf{e}_k\delta V_k$, with random off-diagonal elements describing the response of neighboring QDs. 

Importantly, in both tests the perturbed model, with disorder strength $s$ defined in detail in Appendix~C, is used only to generate the input conductance maps, while the physics decoder retains the ideal Hamiltonian $\mathcal{H}(P)$ of Eq.~\ref{eq:qds} and assumes perfectly local gate control. The resulting tests therefore directly probe the robustness of Hamiltonian inference against model mismatch not represented explicitly in the latent parameterization.

In Fig.~\ref{fig:diss}, the conductance reconstruction error $\langle|G-\hat{G}|\rangle$, averaged over the same parameter ranges as in Fig.~\ref{fig:rashba_hl}(a-c), is shown as a function of the disorder strength $s$ (see Appendix~C). It increases smoothly with $s$, as expected from the increasing mismatch between the Hamiltonian generating the data and the clean model assumed by the decoder. 
%Importantly, the degradation remains gradual rather than abrupt, indicating that the learned inference is robust against moderate deviations from the assumed Hamiltonian family. This also suggests that the additional conductance features generated by the hidden degrees of freedom cannot simply be absorbed into some shifts of the seven inferred Hamiltonian parameters $P$.
The reconstruction error increases continuously with disorder strength, reaching approximately five times the clean baseline for the strongest Hamiltonian perturbations considered. This indicates a progressive loss of reconstruction accuracy under model mismatch, rather than an abrupt failure of the inference procedure. The increasing residual error also suggests that the effects of hidden Hamiltonian perturbations cannot be fully absorbed into shifts of the seven inferred parameters $P$.

\subsection{Beyond the assumed Hamiltonian family}

\begin{figure}[tb]
	\centering
    \includegraphics[width=0.99\columnwidth]{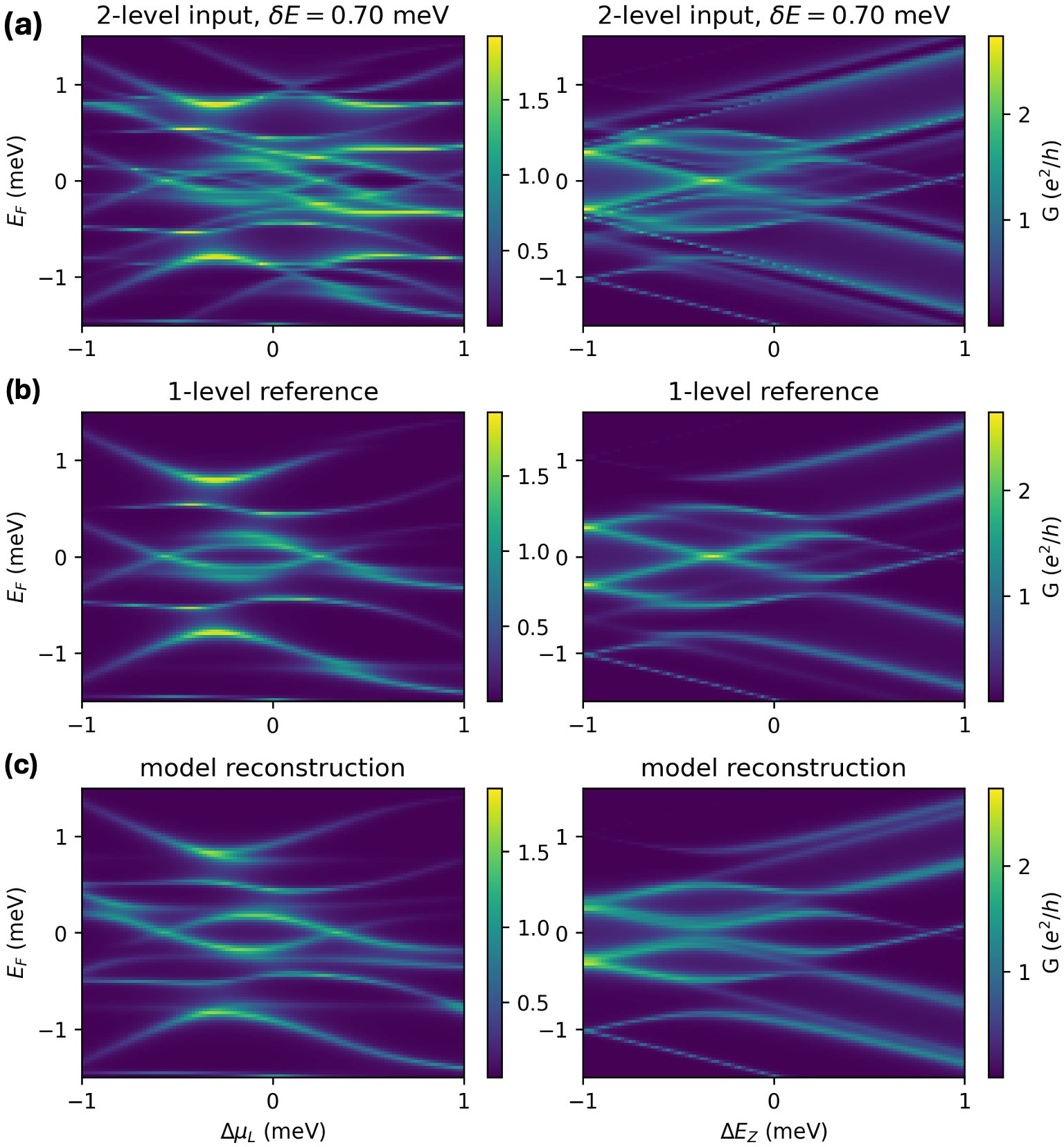}
	\caption{Example conductance maps illustrating model predictions for an extended Hamiltonian with a second-level splitting of $\delta E=0.7$~meV: (a) input maps generated with an additional orbital level on each QD, and (c) their reconstruction by the model retaining the single-level physics decoder; (b) reference generated with the single-level Hamiltonian.}
	\label{fig:two_level_maps}
\end{figure}

We finally consider a more stringent test in which the system generating
the input conductance maps no longer belongs to the Hamiltonian family $\mathcal{H}$
encoded in PD. In a realistic device, additional degrees
of freedom that are absent from the effective model may contribute to
transport. In such a situation, the objective is not necessarily to recover
the full microscopic Hamiltonian, which by construction cannot be
represented by the decoder, but rather to determine whether the measured
conductance can still be described by an effective Hamiltonian within the
assumed model family.
This observation also provides a useful interpretation of the reconstruction
error. A failure to accurately reconstruct conductance maps generated by a Hamiltonian outside the assumed family does not necessarily constitute a
failure of the inference procedure. Instead, a large reconstruction error
can serve as an indicator that the assumed effective Hamiltonian is no
longer adequate to describe the measured system.

To investigate this scenario in a controlled setting, we extend the
Hamiltonian used to generate the input data by introducing a second orbital
level on each QD, while keeping the Hamiltonian implemented in the
PD unchanged. The energy of the additional level is shifted by
$\delta E$ relative to the original level. Thus, $\delta E$ provides a
continuous control parameter for the mismatch between the data-generating
Hamiltonian and the single-level model assumed by the decoder. For large
$\delta E$, the additional level is energetically separated from the
relevant transport window and the original single-level description is
recovered. Upon decreasing $\delta E$, the additional levels enter the
transport window and generate conductance features that cannot, in general,
be represented by the Hamiltonian family used by the decoder. The two-level Hamiltonian is explicitly defined in Appendix~D.

In Fig.~\ref{fig:two_level_maps}(a), we show example conductance maps for a
two-level system with $\delta E=0.7$~meV, while the corresponding reference
single-level maps are shown in Fig.~\ref{fig:two_level_maps}(b). The additional
orbital produces multiple resonant features that are absent from the
corresponding single-level conductance maps.
As can be seen in Fig.~\ref{fig:two_level_maps}(c), when these two-level maps
are provided as input to the network, the predictor does not collapse or produce
an unphysical reconstruction. Instead, it successfully reconstructs a close conductance pattern accessible within its single-level Hamiltonian
family $\mathcal{H}$. This is an encouraging result, as it shows that even when the input
data contain features originating from degrees of freedom absent from the
assumed model, the network can still identify a meaningful effective
Hamiltonian within the model family encoded in the PD.

\begin{figure}[tb]
	\centering
    \includegraphics[width=0.99\columnwidth]{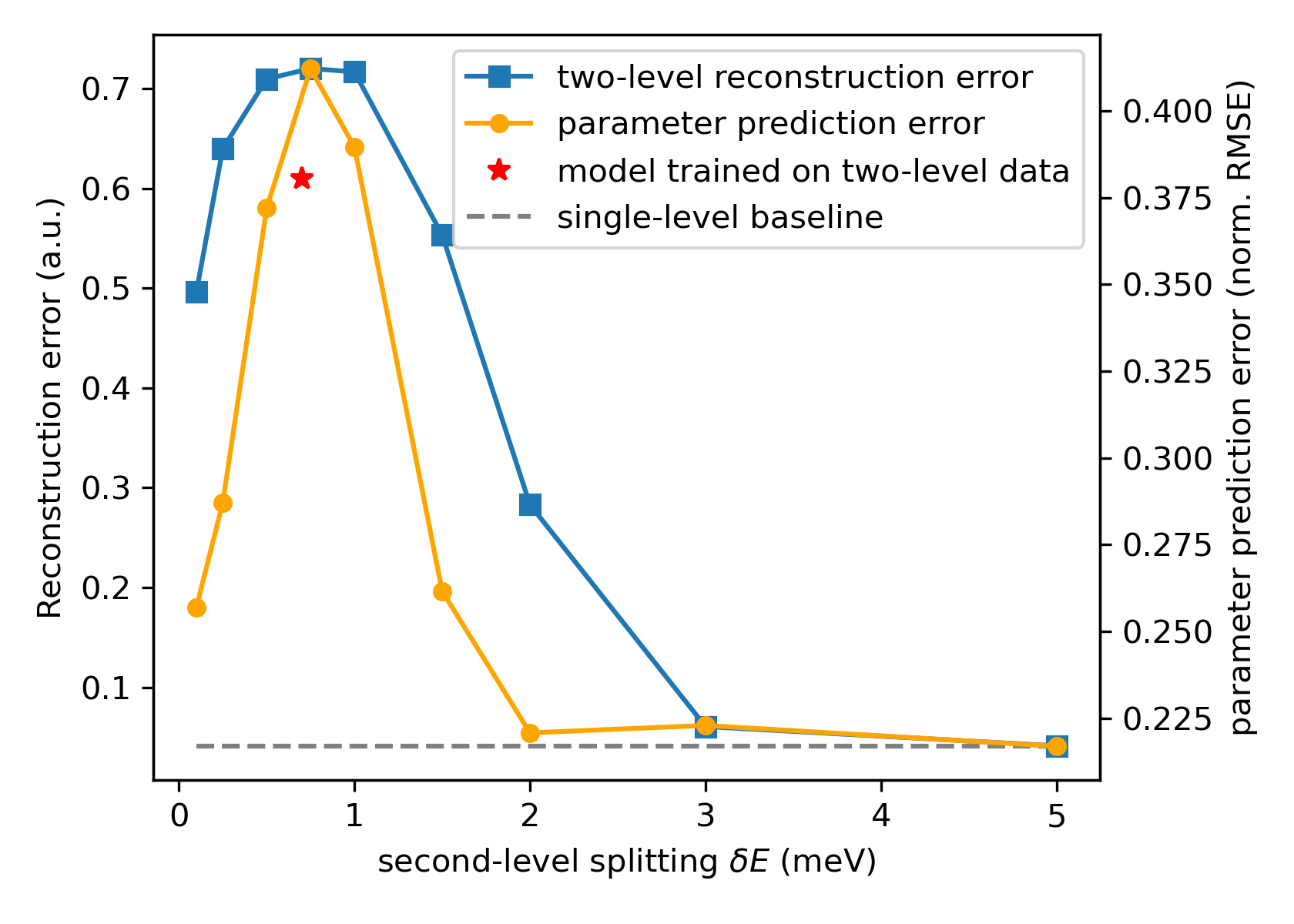}
	\caption{Prediction beyond the $\mathcal{H}$ family assumed in the PD. Reconstruction (blue) and normalized parameter prediction (orange) error for a model trained exclusively on single-level data and subsequently tested on $G$ maps generated with an additional orbital level on each QD, as functions of the second-level splitting $\delta E$.
    The red star shows the reconstruction error for a model trained and tested on two-level data at $\delta E=0.7$~meV, while retaining the single-level $\mathcal{H}$ in the PD.}
	\label{fig:bhf}
\end{figure}

However, the remaining discrepancy between the input and reconstructed maps is
reflected in an increased reconstruction error, presented in Fig.~\ref{fig:bhf}, providing a natural
indicator that the assumed Hamiltonian family is incomplete. As $\delta E$ increases in Fig.~\ref{fig:bhf}, the additional
level becomes effectively decoupled and the reconstruction error approaches the single-level baseline (gray dashed line).
At the same time, the parameter prediction error (orange curve), calculated as the normalized RMSE with respect to the reference single-level parameters $P[\mathcal{H}]$, $(\frac{1}{7}\sum_i(P_i[\mathcal{H}]-\hat{P}_i)^2/{\Delta P_i^2})^{1/2}$, where $\Delta P_i$ denotes the natural range of the $i$-th parameter, increases by only about a factor of two. This indicates that, despite the mismatch with the assumed Hamiltonian family $\mathcal{H}$, the prediction of meaningful effective parameters does not collapse.

We further demonstrate that training directly on two-level data with $\delta E=0.7$~meV is feasible even when retaining the standard single-level Hamiltonian in the PD. The resulting reconstruction error is marked by the red star in Fig.~\ref{fig:bhf}. Interestingly, the model trained locally at a given $\delta E$ achieves a slightly lower reconstruction error than the model trained exclusively on single-level data and subsequently tested on two-level data (blue curve).
However, a substantial residual error remains, reflecting the limitations imposed by the assumed PD Hamiltonian family.

\section{Architecture complexity and scalability}

\subsection{Predictor network complexity}
\begin{figure}[tb]
	\centering
    \includegraphics[width=0.99\columnwidth]{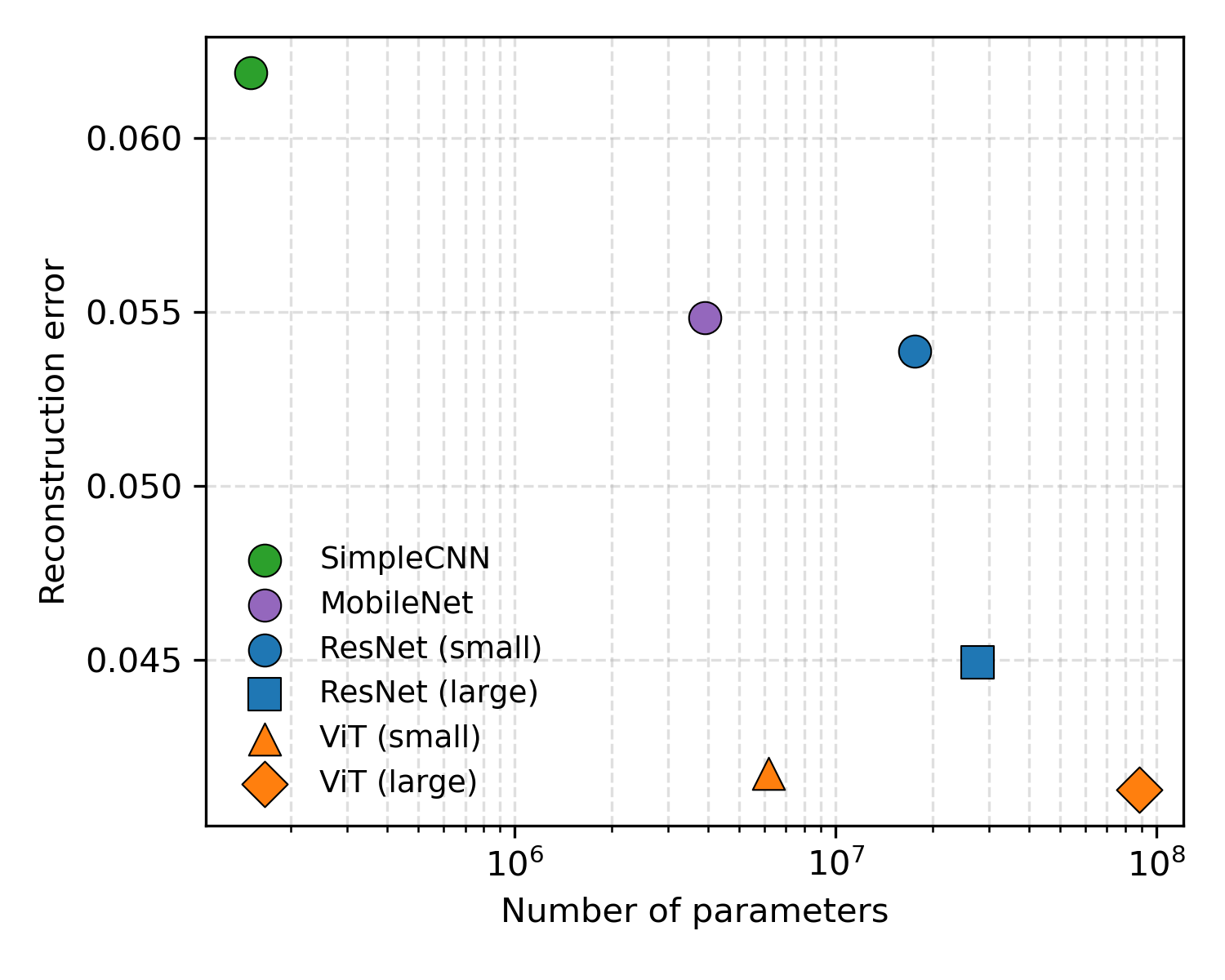}
	\caption{Comparison of various NN architectures serving as the encoder. Predictor complexity is measured by the number of network parameters.}
	\label{fig:other_arch}
\end{figure}

To examine whether the performance of the ViT encoder originates simply from its large number of trainable parameters, we compared several encoder architectures spanning more than two orders of magnitude in model size,
while keeping the physics decoder and training procedure unchanged.
Fig.~\ref{fig:other_arch} shows the reconstruction error as
a function of the number of trainable encoder parameters.

The results indicate that performance depends not only on model size but also strongly on the encoder architecture. A simple CNN yields the largest reconstruction error, while increasing the depth and complexity of residual CNN architectures leads to substantial improvement. Importantly, even the small ViT, with approximately an order of magnitude fewer parameters than the ViT used in the main calculations, outperforms the tested CNN-based architectures of comparable or larger size. This suggests different scaling behavior of transformer- and convolution-based architectures for the present conductance-map inference problem. At the same time, the ViT performance appears to approach saturation with increasing model size, at least for the considered training-set size and corresponding problem complexity.

\subsection{Model scalability}
\begin{figure}[b]
	\centering
    \includegraphics[width=0.8\columnwidth]{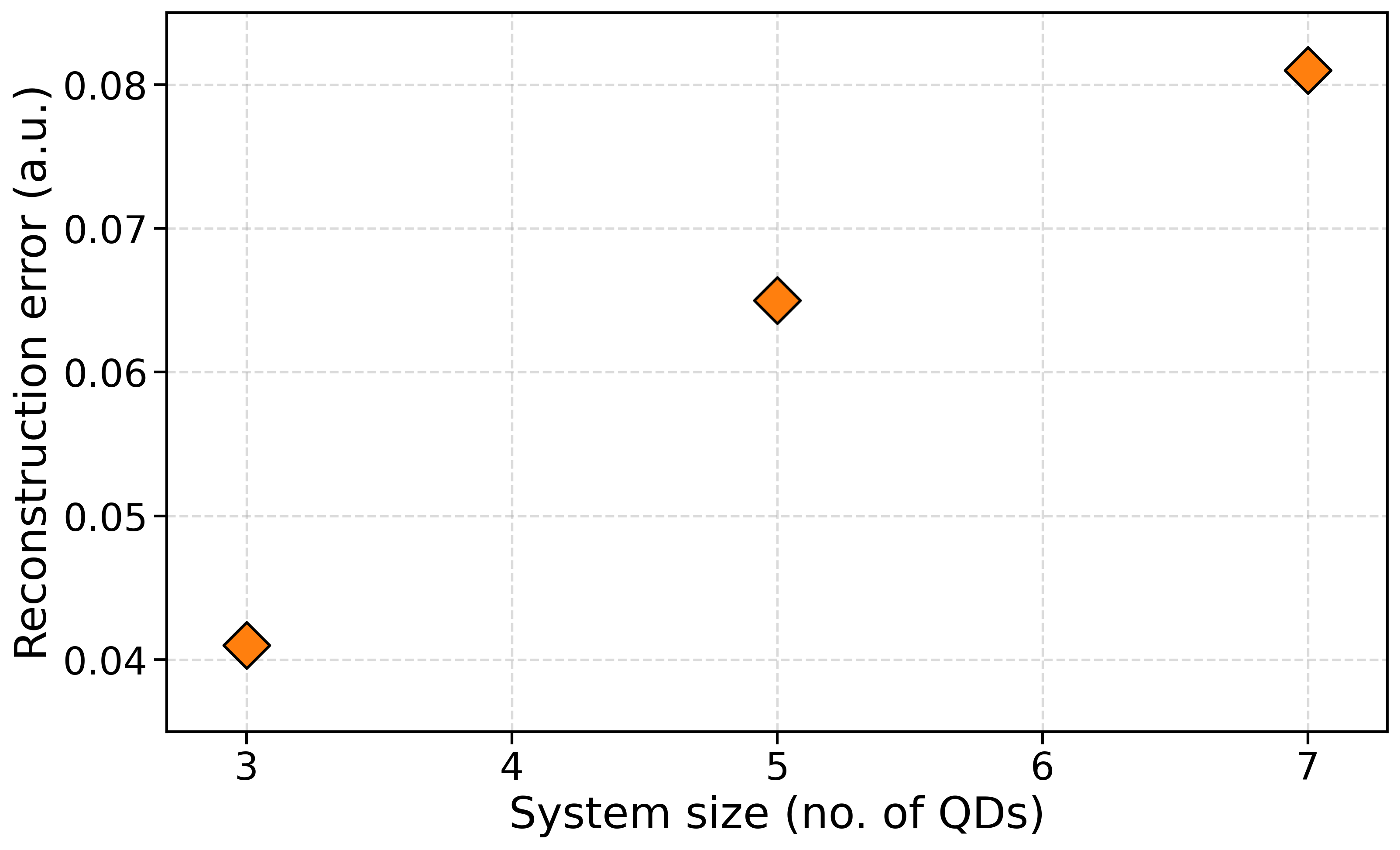}
	\caption{
Scaling with the QD-chain length. Reconstruction error obtained with the
ViT (large) encoder for 3,5,7-QD systems, averaged over the same ranges as in Fig.~\ref{fig:rashba_hl}(a-c).}
	\label{fig:scaling}
\end{figure}
To assess the scalability of the proposed approach, we repeated the same
learning experiment for longer QD chains, extending the system from three
to five and seven QDs. The input and output dimensions were scaled
accordingly, while the ViT (large) encoder architecture and training
procedure were otherwise kept unchanged. Fig.~\ref{fig:scaling} shows the
reconstruction error averaged over the same parameter ranges as in
Fig.~\ref{fig:rashba_hl}(a-c). As expected, the reconstruction error
increases with system size, reflecting the growing complexity of the inverse
problem. The degradation is, however, gradual rather than abrupt: the
five-QD system retains good reconstruction accuracy, while the seven-QD
system shows a larger but still controlled increase in error. These results
indicate that the method remains applicable to longer chains without
architectural modifications, although increasingly large systems are
expected to require additional model capacity and/or training data.

\section{Summary}
In this work we present an architecture and training algorithms for deep neural networks capable of identifying (predicting or discovering---depending on the application context) the parameters of Hamiltonians in quantum systems, including quantum simulators based on nanostructures.
This work focuses on quantum simulators described by lattice Hamiltonians implemented in chains of quantum dots (QDs), together with their basic characterization through transport measurements. Specifically, it addresses the reconstruction of the effective Hamiltonian of a gated QD chain from conductance maps as a function of the applied gate voltages. We propose a natural mechanism for enforcing the underlying physics within the learning procedure.
Moreover, the proposed scheme for defining and training a physics-informed neural network model can be applied to other areas of physics---wherever the underlying equations can be embedded in the form of differentiable formulas.

\section*{Data Availability Statement}

The code used to generate the training and test datasets,
implement the machine-learning models, and perform their
training is publicly available in the GitHub repository~\cite{PawlowskiHlearning}, allowing the datasets
and results presented in this work to be reproduced.

\section*{Acknowledgments}
%MK and JP acknowledge support from National Science Centre, Poland, under grant no. 2021/43/D/ST3/01989. 
We gratefully acknowledge Polish high-performance computing infrastructure PLGrid (HPC Centers: ACK Cyfronet AGH) for providing computer facilities and support within computational grant no. PLG/2025/018433.

\appendix
\section{Appendix A: Chain of $N$ Rashba quantum dots}
\subsection{Analytical considerations}

We start with the Rashba-Zeeman-BCS Hamiltonian, defined in Eq.~1, written in the Nambu spinor representation $\hat{\Psi}_n=(c_{n\uparrow},c_{n\downarrow},c^\dagger_{n\uparrow},c^\dagger_{n\downarrow})^T$ giving $H^\mathrm{QDs}=\frac{1}{2}\sum_{nm}\hat{\Psi}^\dagger_n H^\mathrm{BdG}_{nm}\hat{\Psi}_m$, where $H^\mathrm{BdG}$ is a first-quantized Hamiltonian, also called the Bogoliubov-de Gennes (BdG) Hamiltonian. Its explicit matrix form reads:
\begin{align}\label{eq:hbdg}
&H^\mathrm{BdG}=\nonumber\\
&\sum_{n=1}^{N}
\begin{pmatrix}
-\mu_n+V_\mathrm{Z} & 0 & 0 & \Delta_n\\
0 & -\mu_n-V_\mathrm{Z} & -\Delta_n & 0\\
0 & -\Delta_n & \mu_n-V_\mathrm{Z} & 0\\
\Delta_n & 0 & 0 & \mu_n+V_\mathrm{Z}
\end{pmatrix}_{\!n,n}\nonumber\\
+&\sum_{n=1}^{N-1}
t_n\begin{pmatrix}
e^{i\lambda_n\sigma_y} & 0\\
0 & -e^{-i\lambda_n\sigma_y^\ast}
\end{pmatrix}_{\!n,n+1}+\mathrm{h.c.},
\end{align}
with the onsite (dot) block $(\quad)_{n,n}$ and the offsite hopping (inter-dot) block $(\quad)_{n,n+1}$---both represented by $4\times 4$ matrices. In Eq.~\ref{eq:hbdg} we assumed $\boldsymbol{\lambda}_n \cdot \boldsymbol{\sigma}=\lambda_n\sigma_y$, which follows from choosing the Rashba vector in a form $\boldsymbol\lambda_n=\lambda_n[0,1,0]$.
For a system consisting of $N$ quantum dots, the full BdG Hamiltonian $H^\mathrm{BdG}$ is a $4N\times4N$ matrix.

Our goal is to simulate the Kitaev chain (KC) model, i.e., to tune the $H^\mathrm{BdG}$ (parameters) to mimic KC Hamiltonian form.
The first step is that each
QD should have a pair of zero energy levels $E_n=0$. Diagonalizing a (separated) single dot block $(\quad)_{n,n}$ yields the condition for this to happen: $V_\mathrm{Z}=\sqrt{\Delta^2_n+\mu^2_n}$ or equivalently $\Delta_n=\sqrt{V^2_\mathrm{Z}-\mu^2_n}$.
When this condition is satisfied, each dot has a pair of fermionic excitations $a_n, a^\dagger_n$, where
\begin{equation}
a_n=\frac{1}{\sqrt{2V_\mathrm{Z}}}\left(\sqrt{V_\mathrm{Z}+\mu_n}\,c^\dagger_{n\uparrow}-\sqrt{V_\mathrm{Z}-\mu_n}\,c_{n\downarrow}\right)\!.
\end{equation}
The energy of $a_n$ ($a^\dagger_n$) is zero. Two other modes are separated by the energy $\pm2V_\mathrm{Z}$.
Let's now assume that the hopping is much smaller than the
energy of the excited state, $t\ll2V_\mathrm{Z}$, we may project the Hamiltonian~(\ref{eq:hbdg}) onto the Hilbert space spanned by a set $\{a_1,a^\dagger_1,a_2,a^\dagger_2,\dots,a_N,a^\dagger_N\}$. 
The resulting projected Hamiltonian is:
\begin{equation}\label{eq:bdg_6}
\tilde{H}^\mathrm{BdG}=
\sum_{n=1}^{N-1}
\frac{t_n}{V_\mathrm{Z}}\begin{pmatrix}
-\mu_n\cos{\lambda_n} & \Delta_n\sin{\lambda_n}\\
-\Delta_n\sin{\lambda_n} & \mu_n\cos{\lambda_n}
\end{pmatrix}_{\!n,n+1}+\mathrm{h.c.}.
\end{equation}
The goal is to simulate the KC model, which, e.g. for $N=2$ takes the form (in the same Nambu representation $(a_1,a^\dagger_1,a_2,a^\dagger_2)$ as $\tilde{H}^\mathrm{BdG}$):
\begin{equation}\label{eq:kch_6}
H^\mathrm{KC}=
\begin{pmatrix}
\mu'_1 & 0 & -t'_1 & \Delta'_1\\
0 & -\mu'_1 & -\Delta'_1 & t'_1\\
-t'_1 & -\Delta'_1 & \mu'_2 & 0 \\
\Delta'_1 & t'_1 & 0 & -\mu'_2
\end{pmatrix}\!.
\end{equation}
KC Hamiltonian, Eq.~\ref{eq:kch_6}, at the \emph{sweet spot} $\mu'_n=0$, $t'_n=\Delta'_n$ possesses two MZMs.
For the Rashba–Zeeman
$s$-wave Hamiltonian $\tilde{H}^\mathrm{BdG}$, Eq.~\ref{eq:bdg_6}, to simulate $H^\mathrm{KC}$, Eq.~\ref{eq:kch_6}, at the sweet spot, the following condition must be satisfied by the Rashba length
\begin{align}
\lambda_n=\arctan\!\left(\frac{\mu_n}{\Delta_n}\right)\!,
\end{align}
which together with
$\mu_n=\sqrt{V^2_\mathrm{Z}-\Delta^2_n}$, gives the approximate position of the parameters sweet spot for the QDs Hamiltonian $H^\mathrm{QDs}$.

\subsection{Parameters sweet spot}
If we put the assumed $V_\mathrm{Z}=0.5$~meV and $\Delta_n=0.25$~meV, 
$\sqrt{V^2_\mathrm{Z}-\Delta^2_n}$ gives approximately $0.43$~meV, and $\arctan(\frac{\mu_n}{\Delta_n})\simeq0.33\,\pi$.
However, the numerically tuned sweet spot values for $N=3$ are slightly different: $\mu_n=0.6$~meV and $\lambda_n = 0.27\,\pi$.
These parameters are optimized to ensure that the formed MZMs exhibit both large edge localization and preserved particle-hole symmetry, as shown in Figs.~2(b,c) of the main text. The latter values are selected as the reference parameters, $P_0$ in the main text.

\section{Appendix B: Conductance noise models}
The synthetic conductance maps $G_{ij}(E_F,\Delta l)$ were augmented with two dominant noise contributions according to
\begin{equation}
\tilde{G}_{ij}(E_F,\Delta l)=G_{ij}^{\mathrm{warp}}(E_F,\Delta l)
+\delta G_{ij}^{\mathrm{add}}(E_F,\Delta l),
\end{equation}
where $E_F$ is the Fermi energy in the leads and $l$ denotes the detuned control parameter.

The warping contribution was implemented as a stochastic deformation of the detuning axis,
\begin{equation}
G_{ij}^{\mathrm{warp}}(E_F,\Delta l)
=G_{ij}\big(E_F,\Delta l+\delta l(E_F)\big),
\end{equation}
where $\delta l(E_F)$ represents a random shift of the effective detuning value for each scan line at fixed $E_F$ ($G$ map is composed of 100 lines). 
This shift was composed of a smooth drift component 
$\delta l_{\mathrm{d}}(E_F)$ and a random telegraph component 
$\delta l_{\mathrm{t}}(E_F)$:
\begin{equation}
\delta l(E_F)=\delta l_{\mathrm{d}}(E_F)+\delta l_{\mathrm{t}}(E_F).
\end{equation}
The smooth drift component was generated as a low-frequency random process along the 
scan direction,
\begin{equation}
\delta l_{\mathrm{d}}(E_F)
=\sigma_{\mathrm{d}}\, f(E_F),
\end{equation}
where $f(E_F)$ is a zero-mean Gaussian random function smoothed along the $E_F$ axis, with a correlation length of $5$-$30$ scan lines, meaning that significant changes of $\delta l_{\mathrm{d}}(E_F)$ occur only after several consecutive lines rather than between neighboring conductance pixels; and $\sigma_{\mathrm{d}}$ controls the drift amplitude (randomly sampled between $0.2\%$ and $2\%$ of the detuning range).

The random telegraph component was defined as a piecewise-constant process,
\begin{equation}
\delta l_{\mathrm{t}}(E_F)=\sum_{n=1}^{N_{\mathrm{j}}} A_n\, \Theta(E_F-E_n),
\end{equation}
where $N_{\mathrm{j}}$ is the number of telegraph jumps (randomly sampled between $0$ and $3$), $E_n$ are randomly chosen scan-line positions, $A_n$ are jump amplitudes uniformly sampled between $0.5\%$ and $5\%$ of the detuning range, and $\Theta$ denotes the Heaviside step function. Each term produces a sudden shift of conductance features for 
all subsequent scan lines.

The additive noise contribution was defined as
\begin{equation}
\delta G_{ij}^{\mathrm{add}}(E_F,\Delta l)
=\eta_{\mathrm{w}}(E_F,\Delta l)+\eta_{\mathrm{c}}(E_F,\Delta l),
\end{equation}
where $\eta_{\mathrm{w}}(E_F,\Delta l)$ is an uncorrelated Gaussian white noise and $\eta_{\mathrm{c}}(E_F,\Delta l)$ is a Gaussian noise correlated along the scan direction. 
The standard deviation of $\eta_{\mathrm{w}}$ was randomly sampled between $0.5\%$ and $3\%$ of the maximal conductance value $G_\mathrm{max}$.
The correlated component $\eta_{\mathrm{c}}$ had correlation length between $2$ and $10$ scan lines and amplitude between $0.2\%$ and $2\%$ of $G_\mathrm{max}$. During training, all noise parameters were randomly sampled within these ranges.

\section{Appendix C: Hamiltonian disorder parametrizations}
The Hamiltonian disorder is introduced through static random variations of the local pairing, Zeeman energy, and Rashba spin-orbit orientation: $\Delta_n^{\mathrm{dis}}= \Delta_n \left(1+s\sigma_{\Delta}\xi_n^{\Delta}\right)$,
$V_{\mathrm{Z},n}^{\mathrm{dis}}=V_\mathrm{Z}
\left(1+s\sigma_\mathrm{Z}\xi_n^\mathrm{Z}\right)$, and $(\rho_j^{\mathrm{dis}}= \rho_j+s\sigma_{\mathrm{SO}}\xi_j^\rho, \chi_j^{\mathrm{dis}}= \chi_j+s\sigma_{\mathrm{SO}}\xi_j^\chi)$, with $(\rho,\chi)$ being spherical parameterization of the $\hat{\boldsymbol{\lambda}}$ versor. All $\xi$ are independent random variables drawn from the standard
normal distribution, $\mathcal{N}(0,1)$. We use
$\sigma_{\Delta}=\sigma_\mathrm{Z}=0.1$ and
$\sigma_{\mathrm{SO}}=10^\circ$, while $s\in[0,1]$ controls the overall disorder strength.

As a second test, we account for cross-capacitance between electrostatic
gates. Instead of assuming that detuning gate $k$ modifies only the
corresponding local potential $\mu_k$, we introduce a lever-arm matrix
$\mathbf{A}$ such that
\begin{equation}
    \delta\boldsymbol{\mu}
    =
    \mathbf{A}\mathbf{e}_k\,\delta V_k,
    \qquad
    \mathbf{A}=\mathbf{I}+\mathbf{C},
\end{equation}
where $\mathbf{e}_k$ selects the detuned gate. The diagonal elements
satisfy $C_{ii}=0$, while the off-diagonal lever arms are independently
sampled according to uniform distribution,
\begin{equation}
    C_{ij}\sim\mathcal{U}(-cs,cs),
    \qquad i\neq j.
\end{equation}
We vary the maximum cross-capacitance strength from $0$ to $0.2$, i.e., $c=0.2$ and $s\in[0,1]$.

\section{Appendix D: Multilevel quantum-dot Hamiltonian}

To investigate the robustness of the proposed HL framework beyond the Hamiltonian family assumed in the PD, we extend the single-level QD model of Eq.~\ref{eq:qds} to include multiple orbital levels on each quantum dot. The generalized Hamiltonian takes the form

\[
\begin{aligned}
H_{\mathrm{multi}} = \sum_{n,a,s,s'} \Big[&
\left(-\mu_n+\varepsilon_a\right)\delta_{ss'}
c^\dagger_{n,a,s}c_{n,a,s'}\\
&+V_Z(\sigma_z)_{ss'}c^\dagger_{n,a,s}c_{n,a,s'}
\\
&+\frac{1}{2}\left(
\Delta_n(i\sigma_y)_{ss'}
c^\dagger_{n,a,s}c^\dagger_{n,a,s'}
+\mathrm{h.c.}\right)
\Big]
\\
+\sum_{n,a,s,s'}\Big[&
t_{n,a}\left(e^{i\boldsymbol{\lambda}_{n,a}\cdot\boldsymbol{\sigma}}\right)_{ss'}
c^\dagger_{n,a,s}c_{n+1,a,s'}
+\mathrm{h.c.}
\Big],
\end{aligned}
\]
where \(n\) labels the quantum dots, \(a=1,\ldots,M\) denotes the orbital levels, and \(s,s'\) are spin indices. The orbital energy offsets \(\varepsilon_a\) are measured relative to the lowest orbital, with \(\varepsilon_1=0\). For simplicity, we assume orbital-independent interdot hopping and spin-orbit coupling, i.e., $t_{n,a}=t_n$ and $\boldsymbol{\lambda}_{n,a}=\boldsymbol{\lambda}_{n}$.

For the two-level model considered in Figs.~\ref{fig:two_level_maps} and~\ref{fig:bhf}, we set \(M=2\) and introduce an orbital energy splitting \(\delta E\), such that
\[
\varepsilon_1=0,\qquad \varepsilon_2=\delta E.
\]
As \(\delta E\) increases, the additional orbital moves away from the relevant transport-energy window, and the conductance response approaches that of the single-level model. Conversely, for smaller \(\delta E\), the additional orbital contributes additional resonant transport features that cannot generally be reproduced within the single-level Hamiltonian family. This provides a controlled test of Hamiltonian learning under model mismatch.

\bibliography{bibliography}

@ARTICLE{Leszczynski2024,
  author={Leszczyński, Paweł and Kutorasiński, Kamil and Szewczyk, Marcin and Pawłowski, Jarosław},
  journal={IEEE Transactions on Power Electronics}, 
  title={Machine-Learned Models for Power Magnetic Material Characteristics}, 
  year={2025},
  volume={40},
  number={1},
  pages={1554-1562},
  doi={10.1109/TPEL.2024.3463744},
  url={https://ieeexplore.ieee.org/document/10684126}}

@article{Krawczyk2024a,
  title = {Data-driven criteria for quantum correlations},
  author = {Krawczyk, Mateusz and Paw\l{}owski, Jaros\l{}aw and Ma\ifmmode \acute{s}\else \'{s}\fi{}ka, Maciej M. and Roszak, Katarzyna},
  journal = {Phys. Rev. A},
  volume = {109},
  issue = {2},
  pages = {022405},
  numpages = {12},
  year = {2024},
  month = {Feb},
  publisher = {American Physical Society},
  doi = {10.1103/PhysRevA.109.022405},
  url = {https://link.aps.org/doi/10.1103/PhysRevA.109.022405}
}

@article{Fulga2013,
doi = {10.1088/1367-2630/15/4/045020},
year = {2013},
month = {apr},
publisher = {IOP Publishing},
volume = {15},
number = {4},
pages = {045020},
author = {Fulga, Ion C and Haim, Arbel and Akhmerov, Anton R and Oreg, Yuval},
title = {Adaptive tuning of Majorana fermions in a quantum dot chain},
journal = {New Journal of Physics}
}

@article{Wang2023,
	author = {Dvir, Tom and Wang, Guanzhong and van Loo, Nick and Liu, Chun-Xiao and Mazur, Grzegorz P. and Bordin, Alberto and ten Haaf, Sebastiaan L. D. and Wang, Ji-Yin and van Driel, David and Zatelli, Francesco and Li, Xiang and Malinowski, Filip K. and Gazibegovic, Sasa and Badawy, Ghada and Bakkers, Erik P. A. M. and Wimmer, Michael and Kouwenhoven, Leo P.},
	date = {2023/02/01},
	doi = {10.1038/s41586-022-05585-1},
	id = {Dvir2023},
	isbn = {1476-4687},
	journal = {Nature},
	number = {7948},
	pages = {445--450},
	title = {Realization of a minimal Kitaev chain in coupled quantum dots},
	volume = {614},
	year = {2023}}

@article{bronstein2021,
      title="{Geometric Deep Learning: Grids, Groups, Graphs, Geodesics, and Gauges}", 
      author={Michael M. Bronstein and Joan Bruna and Taco Cohen and Petar Veličković},
      year={2021},
      journal={arXiv preprint arXiv:2104.13478},
url={https://arxiv.org/abs/2104.13478}
}

@article{Leijnse2012,
  title = {Parity qubits and poor man's Majorana bound states in double quantum dots},
  author = {Leijnse, Martin and Flensberg, Karsten},
  journal = {Phys. Rev. B},
  volume = {86},
  issue = {13},
  pages = {134528},
  numpages = {7},
  year = {2012},
  month = {Oct},
  publisher = {American Physical Society},
  doi = {10.1103/PhysRevB.86.134528},
  url = {https://link.aps.org/doi/10.1103/PhysRevB.86.134528}
}

@book{donarini2024,
  title={Quantum Transport in Interacting Nanojunctions},
  author={Donarini, Andrea and Grifoni, Milena},
  year={2024},
  publisher={Springer}
}

@article{Lupi2025,
  title = {Hamiltonian-learning quantum magnets with nonlocal impurity tomography},
  author = {Lupi, Greta and Lado, Jose L.},
  journal = {Phys. Rev. Appl.},
  volume = {23},
  issue = {5},
  pages = {054077},
  numpages = {14},
  year = {2025},
  month = {May},
  publisher = {American Physical Society},
  doi = {10.1103/PhysRevApplied.23.054077},
  url = {https://link.aps.org/doi/10.1103/PhysRevApplied.23.054077}
}

@article{Karjalainen2023,
  title = {Hamiltonian Inference from Dynamical Excitations in Confined Quantum Magnets},
  author = {Karjalainen, Netta and Lippo, Zina and Chen, Guangze and Koch, Rouven and Fumega, Adolfo O. and Lado, Jose L.},
  journal = {Phys. Rev. Appl.},
  volume = {20},
  issue = {2},
  pages = {024054},
  numpages = {11},
  year = {2023},
  month = {Aug},
  publisher = {American Physical Society},
  doi = {10.1103/PhysRevApplied.20.024054},
  url = {https://link.aps.org/doi/10.1103/PhysRevApplied.20.024054}
}

@article{Feng2024,
  title = {Classification of skyrmionic textures and extraction of {H}amiltonian parameters via machine learning},
  author = {Feng, Dushuo and Guan, Zhihao and Wu, Xiaoping and Wu, Yan and Song, Changsheng},
  journal = {Phys. Rev. Appl.},
  volume = {21},
  issue = {3},
  pages = {034009},
  numpages = {11},
  year = {2024},
  month = {Mar},
  publisher = {American Physical Society},
  doi = {10.1103/PhysRevApplied.21.034009},
  url = {https://link.aps.org/doi/10.1103/PhysRevApplied.21.034009}
}

@article{Bonizzoni2022,
  title = {Machine-Learning-Assisted Manipulation and Readout of Molecular Spin Qubits},
  author = {Bonizzoni, Claudio and Tincani, Mirco and Santanni, Fabio and Affronte, Marco},
  journal = {Phys. Rev. Appl.},
  volume = {18},
  issue = {6},
  pages = {064074},
  numpages = {11},
  year = {2022},
  month = {Dec},
  publisher = {American Physical Society},
  doi = {10.1103/PhysRevApplied.18.064074},
  url = {https://link.aps.org/doi/10.1103/PhysRevApplied.18.064074}
}

@article{Darulov2020,
  title = {Autonomous Tuning and Charge-State Detection of Gate-Defined Quantum Dots},
  author = {Darulov\'a, J. and Pauka, S.J. and Wiebe, N. and Chan, K.W. and Gardener, G.C and Manfra, M.J. and Cassidy, M.C. and Troyer, M.},
  journal = {Phys. Rev. Appl.},
  volume = {13},
  issue = {5},
  pages = {054005},
  numpages = {19},
  year = {2020},
  month = {May},
  publisher = {American Physical Society},
  doi = {10.1103/PhysRevApplied.13.054005},
  url = {https://link.aps.org/doi/10.1103/PhysRevApplied.13.054005}
}

@article{Hangleiter2024,
	author = {Hangleiter, Dominik and Roth, Ingo and Fuksa, Jon{\'a}{\v s} and Eisert, Jens and Roushan, Pedram},
	date = {2024/11/06},
	doi = {10.1038/s41467-024-52629-3},
	id = {Hangleiter2024},
	isbn = {2041-1723},
	journal = {Nature Communications},
	number = {1},
	pages = {9595},
	title = {Robustly learning the {H}amiltonian dynamics of a superconducting quantum processor},
	url = {https://doi.org/10.1038/s41467-024-52629-3},
	volume = {15},
	year = {2024}}

@article{Khosravian_2024,
doi = {10.1088/2515-7639/ad1c04},
url = {https://dx.doi.org/10.1088/2515-7639/ad1c04},
year = {2024},
month = {jan},
publisher = {IOP Publishing},
volume = {7},
number = {1},
pages = {015012},
author = {Khosravian, Maryam and Koch, Rouven and Lado, Jose L},
title = {Hamiltonian learning with real-space impurity tomography in topological moiré superconductors},
journal = {Journal of Physics: Materials}
}

@misc{liu2025,
      title={Learning the local density of states of a bilayer moir\'e material in one dimension}, 
      author={Diyi Liu and Alexander B. Watson and Michael Hott and Stephen Carr and Mitchell Luskin},
      year={2025},
      eprint={2405.06688},
      archivePrefix={arXiv},
      primaryClass={math-ph},
      url={https://arxiv.org/abs/2405.06688}, 
}

@article{Lunczer2019,
  title = {Approaching Quantization in Macroscopic Quantum Spin Hall Devices through Gate Training},
  author = {Lunczer, Lukas and Leubner, Philipp and Endres, Martin and M\"uller, Valentin L. and Br\"une, Christoph and Buhmann, Hartmut and Molenkamp, Laurens W.},
  journal = {Phys. Rev. Lett.},
  volume = {123},
  issue = {4},
  pages = {047701},
  numpages = {5},
  year = {2019},
  month = {Jul},
  publisher = {American Physical Society},
  doi = {10.1103/PhysRevLett.123.047701},
  url = {https://link.aps.org/doi/10.1103/PhysRevLett.123.047701}
}

@article{Raissi2019,
title = {Physics-informed neural networks: A deep learning framework for solving forward and inverse problems involving nonlinear partial differential equations},
journal = {Journal of Computational Physics},
volume = {378},
pages = {686-707},
year = {2019},
issn = {0021-9991},
doi = {https://doi.org/10.1016/j.jcp.2018.10.045},
url = {https://www.sciencedirect.com/science/article/pii/S0021999118307125},
author = {M. Raissi and P. Perdikaris and G.E. Karniadakis}
}

@article{Inui2024,
  title = {Inverse {H}amiltonian design of highly entangled quantum systems},
  author = {Inui, Koji and Motome, Yukitoshi},
  journal = {Phys. Rev. Res.},
  volume = {6},
  issue = {3},
  pages = {033080},
  numpages = {15},
  year = {2024},
  month = {Jul},
  publisher = {American Physical Society},
  doi = {10.1103/PhysRevResearch.6.033080},
  url = {https://link.aps.org/doi/10.1103/PhysRevResearch.6.033080}
}

@article{Zhang2018,
  title = {Machine Learning Topological Invariants with Neural Networks},
  author = {Zhang, Pengfei and Shen, Huitao and Zhai, Hui},
  journal = {Phys. Rev. Lett.},
  volume = {120},
  issue = {6},
  pages = {066401},
  numpages = {6},
  year = {2018},
  month = {Feb},
  publisher = {American Physical Society},
  doi = {10.1103/PhysRevLett.120.066401},
  url = {https://link.aps.org/doi/10.1103/PhysRevLett.120.066401}
}

@article{Lew2023,
	author = {Lew, Andrew J. and Jin, Kai and Buehler, Markus J.},
	date = {2023/05/26},
	doi = {10.1038/s41524-023-01036-1},
	id = {Lew2023},
	isbn = {2057-3960},
	journal = {npj Computational Materials},
	number = {1},
	pages = {80},
	title = {Designing architected materials for mechanical compression via simulation, deep learning, and experimentation},
	url = {https://doi.org/10.1038/s41524-023-01036-1},
	volume = {9},
	year = {2023}}

@article{Yunwei2020,
author = {Yunwei Mao  and Qi He  and Xuanhe Zhao },
title = {Designing complex architectured materials with generative adversarial networks},
journal = {Science Advances},
volume = {6},
number = {17},
pages = {eaaz4169},
year = {2020},
doi = {10.1126/sciadv.aaz4169},
URL = {https://www.science.org/doi/abs/10.1126/sciadv.aaz4169},
eprint = {https://www.science.org/doi/pdf/10.1126/sciadv.aaz4169}}

@article{Schutt2018,
    author = {Schütt, K. T. and Sauceda, H. E. and Kindermans, P.-J. and Tkatchenko, A. and Müller, K.-R.},
    title = {SchNet – A deep learning architecture for molecules and materials},
    journal = {The Journal of Chemical Physics},
    volume = {148},
    number = {24},
    pages = {241722},
    year = {2018},
    month = {03},
    issn = {0021-9606},
    doi = {10.1063/1.5019779},
    url = {https://doi.org/10.1063/1.5019779},
    eprint = {https://pubs.aip.org/aip/jcp/article-pdf/doi/10.1063/1.5019779/16655678/241722\_1\_online.pdf},
}

@article{Qi2025,
	author = {Qi, Ji and Ko, Tsz Wai and Wood, Brandon C. and Pham, Tuan Anh and Ong, Shyue Ping},
	date = {2024/02/26},
	doi = {10.1038/s41524-024-01227-4},
	id = {Qi2024},
	isbn = {2057-3960},
	journal = {npj Computational Materials},
	number = {1},
	pages = {43},
	title = {Robust training of machine learning interatomic potentials with dimensionality reduction and stratified sampling},
	url = {https://doi.org/10.1038/s41524-024-01227-4},
	volume = {10},
	year = {2024}}

@article{Nieuwenburg2017,
	author = {van Nieuwenburg, Evert P. L. and Liu, Ye-Hua and Huber, Sebastian D.},
	date = {2017/05/01},
	doi = {10.1038/nphys4037},
	id = {van Nieuwenburg2017},
	isbn = {1745-2481},
	journal = {Nature Physics},
	number = {5},
	pages = {435--439},
	title = {Learning phase transitions by confusion},
	url = {https://doi.org/10.1038/nphys4037},
	volume = {13},
	year = {2017}}

@article{Merchant2023,
	author = {Merchant, Amil and Batzner, Simon and Schoenholz, Samuel S. and Aykol, Muratahan and Cheon, Gowoon and Cubuk, Ekin Dogus},
	date = {2023/12/01},
	doi = {10.1038/s41586-023-06735-9},
	id = {Merchant2023},
	isbn = {1476-4687},
	journal = {Nature},
	number = {7990},
	pages = {80--85},
	title = {Scaling deep learning for materials discovery},
	url = {https://doi.org/10.1038/s41586-023-06735-9},
	volume = {624},
	year = {2023}}

@article{Cheetham2024,
	author = {Cheetham, Anthony K. and Seshadri, Ram},
	date = {2024/04/23},
	doi = {10.1021/acs.chemmater.4c00643},
	isbn = {0897-4756},
	journal = {Chemistry of Materials},
	journal1 = {Chemistry of Materials},
	journal2 = {Chem. Mater.},
	month = {04},
	number = {8},
	pages = {3490--3495},
	publisher = {American Chemical Society},
	title = {Artificial Intelligence Driving Materials Discovery? Perspective on the Article: Scaling Deep Learning for Materials Discovery},
	type = {doi: 10.1021/acs.chemmater.4c00643},
	url = {https://doi.org/10.1021/acs.chemmater.4c00643},
	volume = {36},
	year = {2024},
	year1 = {2024}}

@article{Gu2024,
	author = {Gu, Qiangqiang and Zhouyin, Zhanghao and Pandey, Shishir Kumar and Zhang, Peng and Zhang, Linfeng and E, Weinan},
	date = {2024/08/08},
	doi = {10.1038/s41467-024-51006-4},
	id = {Gu2024},
	isbn = {2041-1723},
	journal = {Nature Communications},
	number = {1},
	pages = {6772},
	title = {Deep learning tight-binding approach for large-scale electronic simulations at finite temperatures with ab initio accuracy},
	url = {https://doi.org/10.1038/s41467-024-51006-4},
	volume = {15},
	year = {2024}}

@article{Henderson2023,
  title = {Deep learning extraction of band structure parameters from density of states: A case study on trilayer graphene},
  author = {Henderson, Paul and Ghazaryan, Areg and Zibrov, Alexander A. and Young, Andrea F. and Serbyn, Maksym},
  journal = {Phys. Rev. B},
  volume = {108},
  issue = {12},
  pages = {125411},
  numpages = {11},
  year = {2023},
  month = {Sep},
  publisher = {American Physical Society},
  doi = {10.1103/PhysRevB.108.125411},
  url = {https://link.aps.org/doi/10.1103/PhysRevB.108.125411}
}

@article{Mirani2024,
  title = {Learning interacting fermionic {H}amiltonians at the {H}eisenberg limit},
  author = {Mirani, Arjun and Hayden, Patrick},
  journal = {Phys. Rev. A},
  volume = {110},
  issue = {6},
  pages = {062421},
  numpages = {14},
  year = {2024},
  month = {Dec},
  publisher = {American Physical Society},
  doi = {10.1103/PhysRevA.110.062421},
  url = {https://link.aps.org/doi/10.1103/PhysRevA.110.062421}
}

@Article{StilckFranca2024,
author={Stilck Fran{\c{c}}a, Daniel
and Markovich, Liubov A.
and Dobrovitski, V. V.
and Werner, Albert H.
and Borregaard, Johannes},
title={Efficient and robust estimation of many-qubit {H}amiltonians},
journal={Nature Communications},
year={2024},
month={Jan},
day={08},
volume={15},
number={1},
pages={311},
issn={2041-1723},
doi={10.1038/s41467-023-44012-5},
url={https://doi.org/10.1038/s41467-023-44012-5}
}

@article{Kovachki2023,
author = {Kovachki, Nikola and Li, Zongyi and Liu, Burigede and Azizzadenesheli, Kamyar and Bhattacharya, Kaushik and Stuart, Andrew and Anandkumar, Anima},
title = {Neural operator: learning maps between function spaces with applications to PDEs},
year = {2023},
issue_date = {January 2023},
publisher = {JMLR.org},
volume = {24},
number = {1},
issn = {1532-4435},
journal = {J. Mach. Learn. Res.},
month = jan,
articleno = {89},
numpages = {97}
}

@article{Li2024,
author = {Li, Zongyi and Zheng, Hongkai and Kovachki, Nikola and Jin, David and Chen, Haoxuan and Liu, Burigede and Azizzadenesheli, Kamyar and Anandkumar, Anima},
title = {Physics-Informed Neural Operator for Learning Partial Differential Equations},
year = {2024},
issue_date = {September 2024},
publisher = {Association for Computing Machinery},
address = {New York, NY, USA},
volume = {1},
number = {3},
url = {https://doi.org/10.1145/3648506},
doi = {10.1145/3648506},
journal = {ACM / IMS J. Data Sci.},
month = may,
articleno = {9},
numpages = {27}
}

@article{Christiansen2009,
  title={A mathematical formulation of the {Mahaux--Weidenm{\"u}ller} formula for the scattering matrix},
  author={Christiansen, TJ and Zworski, M},
  journal={Journal of Physics A: Mathematical and Theoretical},
  volume={42},
  number={41},
  pages={415202},
  year={2009},
  publisher={IOP Publishing}
}

@article{Blonder1982,
  title = {Transition from metallic to tunneling regimes in superconducting microconstrictions: Excess current, charge imbalance, and supercurrent conversion},
  author = {Blonder, G. E. and Tinkham, M. and Klapwijk, T. M.},
  journal = {Phys. Rev. B},
  volume = {25},
  issue = {7},
  pages = {4515--4532},
  numpages = {0},
  year = {1982},
  month = {Apr},
  publisher = {American Physical Society},
  doi = {10.1103/PhysRevB.25.4515},
  url = {https://link.aps.org/doi/10.1103/PhysRevB.25.4515}
}

@article{grepkowa2023,
  title = {Adversarial {H}amiltonian learning of quantum dots in a minimal {Kitaev} chain},
  author = {Koch, Rouven and van Driel, David and Bordin, Alberto and Lado, Jose L. and Greplova, Eliska},
  journal = {Phys. Rev. Appl.},
  volume = {20},
  issue = {4},
  pages = {044081},
  numpages = {13},
  year = {2023},
  month = {Oct},
  publisher = {American Physical Society},
  doi = {10.1103/PhysRevApplied.20.044081}
}

@article{Gebhart2023,
	author = {Gebhart, Valentin and Santagati, Raffaele and Gentile, Antonio Andrea and Gauger, Erik M. and Craig, David and Ares, Natalia and Banchi, Leonardo and Marquardt, Florian and Pezz{\`e}, Luca and Bonato, Cristian},
	date = {2023/03/01},
	doi = {10.1038/s42254-022-00552-1},
	id = {Gebhart2023},
	isbn = {2522-5820},
	journal = {Nature Reviews Physics},
	number = {3},
	pages = {141--156},
	title = {Learning quantum systems},
	volume = {5},
	year = {2023}
}

@article{Wang2017,
	author = {Wang, Jianwei and Paesani, Stefano and Santagati, Raffaele and Knauer, Sebastian and Gentile, Antonio A. and Wiebe, Nathan and Petruzzella, Maurangelo and O'Brien, Jeremy L. and Rarity, John G. and Laing, Anthony and Thompson, Mark G.},
	date = {2017/06/01},
	doi = {10.1038/nphys4074},
	id = {Wang2017},
	isbn = {1745-2481},
	journal = {Nature Physics},
	number = {6},
	pages = {551--555},
	title = {Experimental quantum {H}amiltonian learning},
	volume = {13},
	year = {2017}
}

@article{Zwolak2023,
  title = {Colloquium: Advances in automation of quantum dot devices control},
  author = {Zwolak, Justyna P. and Taylor, Jacob M.},
  journal = {Rev. Mod. Phys.},
  volume = {95},
  issue = {1},
  pages = {011006},
  numpages = {20},
  year = {2023},
  month = {Feb},
  publisher = {American Physical Society},
  doi = {10.1103/RevModPhys.95.011006},
  url = {https://link.aps.org/doi/10.1103/RevModPhys.95.011006}
}

@misc{roux2025,
      title={Rapid Autotuning of a SiGe Quantum Dot into the Single-Electron Regime with Machine Learning and {RF}-Reflectometry {FPGA}-Based Measurements}, 
      author={Marc-Antoine Roux and Joffrey Rivard and Victor Yon and Alexis Morel and Dominic Leclerc and Claude Rohrbacher and El Bachir Ndiaye and Felice Francesco Tafuri and Brendan Bono and Stefan Kubicek and Roger Loo and Yosuke Shimura and Julien Jussot and Clément Godfrin and Danny Wan and Kristiaan De Greve and Marc-André Tétrault and Dominique Drouin and Christian Lupien and Michel Pioro-Ladrière and Eva Dupont-Ferrier},
      year={2025},
      eprint={2509.19537},
      archivePrefix={arXiv},
      primaryClass={cond-mat.mes-hall},
      url={https://arxiv.org/abs/2509.19537}, 
}

@inproceedings{
dosovitskiy2021,
title={An Image is Worth 16x16 Words: Transformers for Image Recognition at Scale},
author={Alexey Dosovitskiy and Lucas Beyer and Alexander Kolesnikov and Dirk Weissenborn and Xiaohua Zhai and Thomas Unterthiner and Mostafa Dehghani and Matthias Minderer and Georg Heigold and Sylvain Gelly and Jakob Uszkoreit and Neil Houlsby},
booktitle={International Conference on Learning Representations},
year={2021},
url={https://openreview.net/forum?id=YicbFdNTTy}
}

@article{Bordin2025,
  author       = {Alberto Bordin and Chun-Xiao Liu and Tom Dvir and Francesco Zatelli and 
                  Sebastiaan L. D. ten Haaf and David van Driel and Guanzhong Wang and 
                  Nick van Loo and Yining Zhang and Jan Cornelis Wolff and 
                  Thomas Van Caekenberghe and Ghada Badawy and Sasa Gazibegovic and 
                  Erik P. A. M. Bakkers and Michael Wimmer and Leo P. Kouwenhoven and 
                  Grzegorz P. Mazur},
  title        = {Enhanced {Majorana} stability in a three-site {Kitaev} chain},
  journal      = {Nature Nanotechnology},
  year         = {2025},
  volume       = {20},
  number       = {6},
  pages        = {726--731},
  doi          = {10.1038/s41565-025-01894-4},
  url          = {https://doi.org/10.1038/s41565-025-01894-4},
  issn         = {1748-3395}
}

@article{Taylor2024disorderlearning,
  title = {Machine Learning the Disorder Landscape of {Majorana} Nanowires},
  author = {Taylor, Jacob R. and Sau, Jay D. and Das Sarma, Sankar},
  journal = {Phys. Rev. Lett.},
  volume = {132},
  issue = {20},
  pages = {206602},
  numpages = {5},
  year = {2024},
  month = {May},
  publisher = {American Physical Society},
  doi = {10.1103/PhysRevLett.132.206602},
  url = {https://link.aps.org/doi/10.1103/PhysRevLett.132.206602}
}

@article{Taylor2025disordermitigation,
  title = {Mitigating disorder and optimizing topological indicators with vision-transformer-based neural networks in {Majorana} nanowires},
  author = {Taylor, Jacob R. and Das Sarma, Sankar},
  journal = {Phys. Rev. B},
  volume = {112},
  issue = {4},
  pages = {L041110},
  numpages = {6},
  year = {2025},
  month = {Jul},
  publisher = {American Physical Society},
  doi = {10.1103/8p7r-cw9k},
  url = {https://link.aps.org/doi/10.1103/8p7r-cw9k}
}

@article{Taylor2025analysis,
  title = {Neural network based deep learning analysis of semiconductor quantum dot qubits for automated control},
  author = {Taylor, Jacob R. and Das Sarma, Sankar},
  journal = {Phys. Rev. B},
  volume = {111},
  issue = {3},
  pages = {035301},
  numpages = {25},
  year = {2025},
  month = {Jan},
  publisher = {American Physical Society},
  doi = {10.1103/PhysRevB.111.035301},
  url = {https://link.aps.org/doi/10.1103/PhysRevB.111.035301}
}

@article{Thamm2024,
  title = {Conductance based machine learning of optimal gate voltages for disordered {Majorana} wires},
  author = {Thamm, Matthias and Rosenow, Bernd},
  journal = {Phys. Rev. B},
  volume = {109},
  issue = {4},
  pages = {045132},
  numpages = {14},
  year = {2024},
  month = {Jan},
  publisher = {American Physical Society},
  doi = {10.1103/PhysRevB.109.045132},
  url = {https://link.aps.org/doi/10.1103/PhysRevB.109.045132}
}

@article{Borsoi2024,
	author = {Borsoi, Francesco and Hendrickx, Nico W. and John, Valentin and Meyer, Marcel and Motz, Sayr and van Riggelen, Floor and Sammak, Amir and de Snoo, Sander L. and Scappucci, Giordano and Veldhorst, Menno},
	date = {2024/01/01},
	doi = {10.1038/s41565-023-01491-3},
	id = {Borsoi2024},
	isbn = {1748-3395},
	journal = {Nature Nanotechnology},
	number = {1},
	pages = {21--27},
	title = {Shared control of a 16 semiconductor quantum dot crossbar array},
	url = {https://doi.org/10.1038/s41565-023-01491-3},
	volume = {19},
	year = {2024}}

@article{Mills2019,
	author = {Mills, A. R. and Zajac, D. M. and Gullans, M. J. and Schupp, F. J. and Hazard, T. M. and Petta, J. R.},
	date = {2019/03/05},
	doi = {10.1038/s41467-019-08970-z},
	id = {Mills2019},
	isbn = {2041-1723},
	journal = {Nature Communications},
	number = {1},
	pages = {1063},
	title = {Shuttling a single charge across a one-dimensional array of silicon quantum dots},
	url = {https://doi.org/10.1038/s41467-019-08970-z},
	volume = {10},
	year = {2019}}

@article{Shandilya2025,
	author = {Shandilya, Aakash and Kapila, Sundeep and Krishnan, Radha and Weber, Bent and Muralidharan, Bhaskaran},
	date = {2025/08/01},
	doi = {10.1021/acsanm.5c01655},
	journal = {ACS Applied Nano Materials},
	journal1 = {ACS Applied Nano Materials},
	journal2 = {ACS Appl. Nano Mater.},
	month = {08},
	number = {30},
	pages = {14949--14959},
	publisher = {American Chemical Society},
	title = {Unified Simulation Framework for Experimentally-Observed Spin-Valley Locking in {MoS2} Quantum Dots: Implications for Qubit Applications},
	type = {doi: 10.1021/acsanm.5c01655},
	url = {https://doi.org/10.1021/acsanm.5c01655},
	volume = {8},
	year = {2025},
	year1 = {2025}}

@article{Maska2017,
	author = {Ma{\'s}ka, Maciej M. and Doma{\'n}ski, Tadeusz},
	date = {2017/11/23},
	doi = {10.1038/s41598-017-16323-3},
	id = {Ma{\'s}ka2017},
	isbn = {2045-2322},
	journal = {Scientific Reports},
	number = {1},
	pages = {16193},
	title = {Polarization of the {Majorana} quasiparticles in the {Rashba} chain},
	url = {https://doi.org/10.1038/s41598-017-16323-3},
	volume = {7},
	year = {2017}}

@misc{vandriel2024,
      title={Cross-Platform Autonomous Control of Minimal Kitaev Chains}, 
      author={David van Driel and Rouven Koch and Vincent P. M. Sietses and Sebastiaan L. D. ten Haaf and Chun-Xiao Liu and Francesco Zatelli and Bart Roovers and Alberto Bordin and Nick van Loo and Guanzhong Wang and Jan Cornelis Wolff and Grzegorz P. Mazur and Tom Dvir and Ivan Kulesh and Qingzhen Wang and A. Mert Bozkurt and Sasa Gazibegovic and Ghada Badawy and Erik P. A. M. Bakkers and Michael Wimmer and Srijit Goswami and Jose L. Lado and Leo P. Kouwenhoven and Eliska Greplova},
      year={2024},
      eprint={2405.04596},
      archivePrefix={arXiv},
      primaryClass={cond-mat.mes-hall},
      url={https://arxiv.org/abs/2405.04596}, 
}

@misc{losert2025,
      title={Automated electrostatic characterization of quantum dot devices in single- and bilayer heterostructures}, 
      author={Merritt P. R. Losert and Dario Denora and Barnaby van Straaten and Michael Chan and Stefan D. Oosterhout and Lucas Stehouwer and Giordano Scappucci and Menno Veldhorst and Justyna P. Zwolak},
      year={2025},
      eprint={2601.00067},
      archivePrefix={arXiv},
      primaryClass={cond-mat.mes-hall},
      url={https://arxiv.org/abs/2601.00067}, 
}

@Article{Choudhary2025,
author={Choudhary, Kamal},
title={SlaKoNet: A Unified Slater-Koster Tight-Binding Framework Using Neural Network Infrastructure for the Periodic Table},
journal={The Journal of Physical Chemistry Letters},
year={2025},
month={Oct},
day={30},
publisher={American Chemical Society},
volume={16},
number={43},
pages={11109-11119},
doi={10.1021/acs.jpclett.5c02456},
url={https://doi.org/10.1021/acs.jpclett.5c02456}
}

@misc{krawczyk2026,
      title={{AI}-enhanced tuning of quantum dot {H}amiltonians toward Majorana modes}, 
      author={Mateusz Krawczyk and Jarosław Pawłowski},
      year={2026},
      eprint={2601.02149},
      archivePrefix={arXiv},
      primaryClass={cond-mat.mes-hall},
      url={https://arxiv.org/abs/2601.02149}, 
}

@article{Kliczkowski2024,
  title = {Autoencoder-based analytic continuation method for strongly correlated quantum systems},
  author = {Kliczkowski, Maksymilian and Keyes, Lauren and Roy, Sayantan and Paiva, Thereza and Randeria, Mohit and Trivedi, Nandini and Ma\ifmmode \acute{s}\else \'{s}\fi{}ka, Maciej M.},
  journal = {Phys. Rev. B},
  volume = {110},
  issue = {11},
  pages = {115119},
  numpages = {15},
  year = {2024},
  month = {Sep},
  publisher = {American Physical Society},
  doi = {10.1103/PhysRevB.110.115119},
  url = {https://link.aps.org/doi/10.1103/PhysRevB.110.115119}
}

@article{Chng2017,
  title = {Machine Learning Phases of Strongly Correlated Fermions},
  author = {Ch'ng, Kelvin and Carrasquilla, Juan and Melko, Roger G. and Khatami, Ehsan},
  journal = {Phys. Rev. X},
  volume = {7},
  issue = {3},
  pages = {031038},
  numpages = {9},
  year = {2017},
  month = {Aug},
  publisher = {American Physical Society},
  doi = {10.1103/PhysRevX.7.031038},
  url = {https://link.aps.org/doi/10.1103/PhysRevX.7.031038}
}

@article{greydanus2019hamiltonian,
  title={Hamiltonian neural networks},
  author={Greydanus, Samuel and Dzamba, Misko and Yosinski, Jason},
  journal={Advances in neural information processing systems},
  volume={32},
  year={2019}
}

@Article{Li2022,
author={Li, He
and Wang, Zun
and Zou, Nianlong
and Ye, Meng
and Xu, Runzhang
and Gong, Xiaoxun
and Duan, Wenhui
and Xu, Yong},
title={Deep-learning density functional theory Hamiltonian for efficient ab initio electronic-structure calculation},
journal={Nature Computational Science},
year={2022},
month={Jun},
day={01},
volume={2},
number={6},
pages={367-377},
issn={2662-8457},
doi={10.1038/s43588-022-00265-6},
url={https://doi.org/10.1038/s43588-022-00265-6}
}

@article{Elhamod2022,
author = {Elhamod, Mohannad and Bu, Jie and Singh, Christopher and Redell, Matthew and Ghosh, Abantika and Podolskiy, Viktor and Lee, Wei-Cheng and Karpatne, Anuj},
title = {
CoPhy-PGNN: Learning Physics-guided Neural Networks with Competing Loss Functions for Solving Eigenvalue Problems},
year = {2022},
issue_date = {December 2022},
publisher = {Association for Computing Machinery},
address = {New York, NY, USA},
volume = {13},
number = {6},
issn = {2157-6904},
url = {https://doi.org/10.1145/3530911},
doi = {10.1145/3530911},
journal = {ACM Trans. Intell. Syst. Technol.},
month = dec,
articleno = {92},
numpages = {23}
}

@article{Mattheakis2022,
  title = {Hamiltonian neural networks for solving equations of motion},
  author = {Mattheakis, Marios and Sondak, David and Dogra, Akshunna S. and Protopapas, Pavlos},
  journal = {Phys. Rev. E},
  volume = {105},
  issue = {6},
  pages = {065305},
  numpages = {11},
  year = {2022},
  month = {Jun},
  publisher = {American Physical Society},
  doi = {10.1103/PhysRevE.105.065305},
  url = {https://link.aps.org/doi/10.1103/PhysRevE.105.065305}
}

@misc{hernandes2026,
      title={Reconstructing Quantum Dot Charge Stability Diagrams with Diffusion Models}, 
      author={Vinicius Hernandes and Joseph Rogers and Rouven Koch and Thomas Spriggs and Brennan Undseth and Anasua Chatterjee and Lieven M. K. Vandersypen and Eliska Greplova},
      year={2026},
      eprint={2603.26432},
      archivePrefix={arXiv},
      primaryClass={quant-ph},
      url={https://arxiv.org/abs/2603.26432}, 
}

@misc{PawlowskiHlearning,
  author       = {Pawlowski, J. and Krawczyk, M.},
  title        = {Hlearning: Code for Hamiltonian learning
                  in quantum-dot systems},
  howpublished = {\url{https://github.com/jarek-pawlowski/Hlearning}},
  year         = {2026},
  note         = {{GitHub} repository}
}

\end{document}